# Biblio-Analysis of Cohort Intelligence (CI) Algorithm and its allied applications from Scopus and Web of Science Perspective

Ishaan Kale*[1], Rahul Joshi[2], Kalyani Kadam[3]

Symbiosis Institute of Technology (SIT), Symbiosis International (Deemed University), Pune-412115, India

*ishaan.kale@sitpune.edu.in[1], rahulj@sitpune.edu.in[2], kalyanika@sitpune.edu.in[3]*

**ABSTRACT**

Cohort Intelligence or CI is one of its kind of novel optimization algorithm. Since its inception, in a very short span it is applied successfully in various domains and its results are observed to be effectual in contrast to algorithm of its kind. Till date, there is no such type of bibliometric analysis carried out on CI and its related applications. So, this research paper in a way will be an ice breaker for those who want to take up CI to a new level. In this research papers, CI publications available in Scopus are analyzed through graphs, networked diagrams about authors, source titles, keywords over the years, journals over the time. In a way this bibliometric paper showcase CI, its applications and detail outs systematic review in terms its bibliometric details.

**1. Introduction**

The problems from engineering domain mainly involve mixed and discrete variables design, and nonlinear and linear constraints. This increases the complexity of problems. The traditional optimization techniques may not be suitable to get the feasible solution or unable to handle such problems. So various heuristic and metaheuristic algorithms were proposed by the researchers so far and employed to solve complex problems. These techniques are generally inspired by nature. They are broadly classified into sub-domains such as bio-inspired, socio-inspired and physics & chemical-inspired optimization techniques Fig. 1 [1].

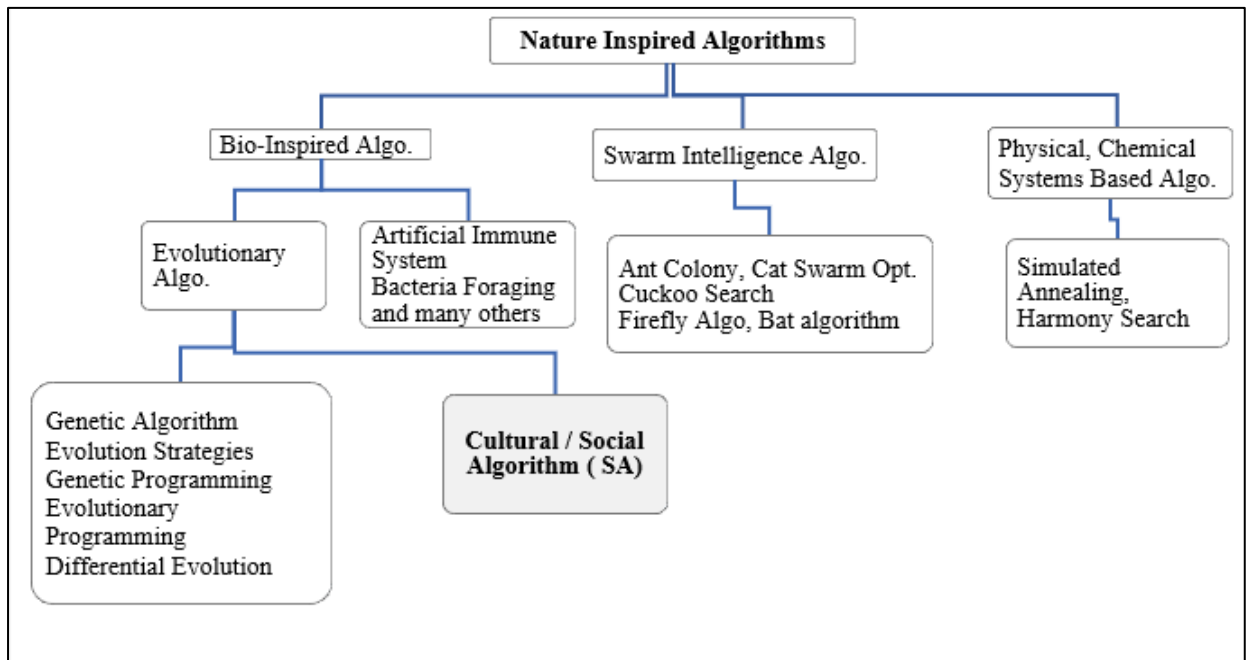

**Fig. 1 Classification of Nature Inspired Optimization Algorithms [1]**

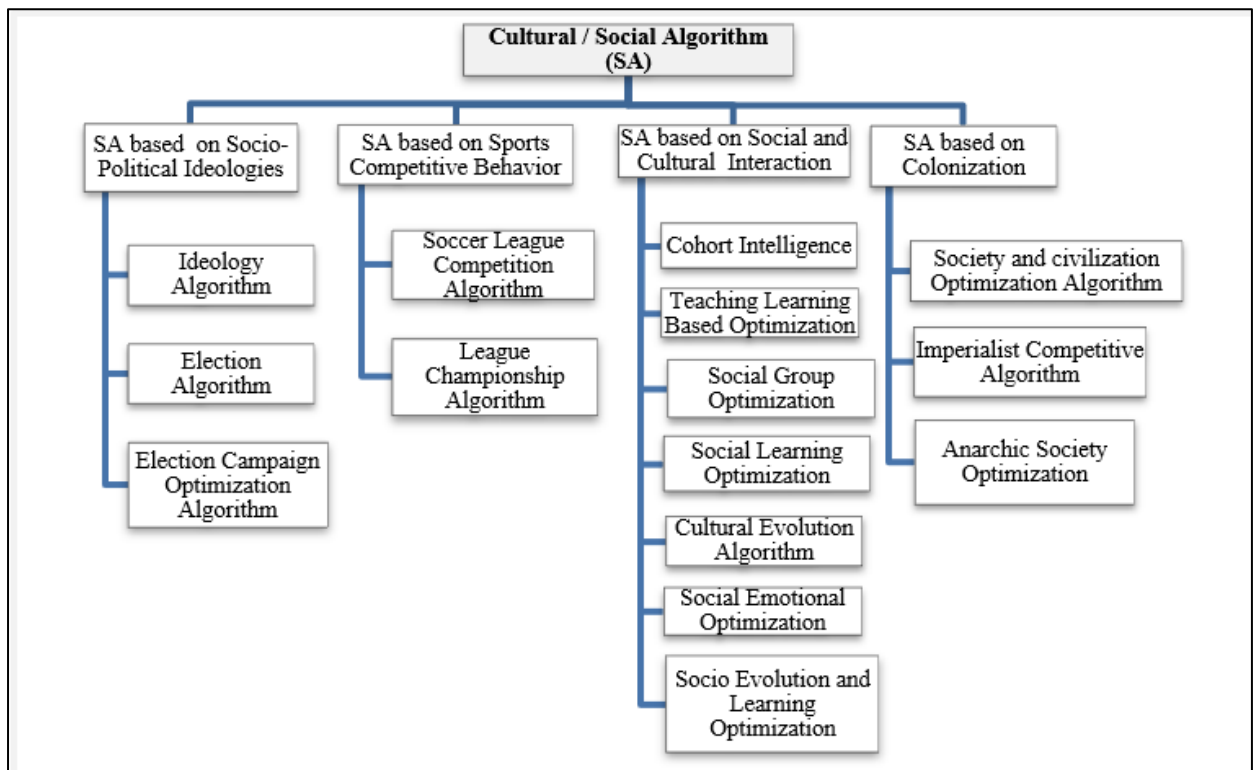

**Fig. 2 Classification of Cultural and Social Algorithms [1, 2]**

Cohort Intelligence (CI) a socio-oriented algorithm was tested and investigated in the wide range of engineering applications as well as on different data sets on health care and clustering problems

[1-5]. The bibliometric survey on metaheuristic CI algorithm could be very much beneficial to new researchers to evolve the algorithm using different natural and philosophical laws and its applicability. This work also exhibits the detailed information about keywords and publication to make the search easier to find the article on a specific application. The detailed illustration of CI version along with its applications is presented in Table 1.

An Artificial Intelligence driven inspired by social tendencies is a Cohort Intelligence (CI) algorithm [7-8]. The tendencies exhibited by social behavior triggers learning of candidates within the cohort. In the cohort every candidate recursively tries to achieve a goal which is in a way common to everyone [6]. For this, roulette wheel approach is employed by each of them, and selection of another candidate is done resulting into own behavior improvement. Learning is triggered through learning achieved from one-another and the entire cohort behavior get evolved. The behavior of the cohort gets saturated, after considerable number of learning attempts if there is no improvement in the behavior of candidates and the cohort behavior is almost the stable or saturated [9-11]. The flowchart is present in Fig. 3. The CI algorithm characteristics are as follows:

- Learning of candidates in the cohort gets modeled. Candidates have Inherently common goal leading to achieve the best behavior by improving its qualities. The interaction and competition are the two natural instincts of every cohort individual. These are achieved through roulette wheel selection and further sampling in the close neighborhood of the selected (being followed) candidate. For details refer to [13-15].
- Improvement in individual behavior and related qualities is through by every candidate's observation and through observation related to every other candidate in the cohort.
- Search space gets updated at the end of each learning attempt; also, each candidate updates its search space by its own.

- Large number constraints and variable problems are effectively handled [1-7].

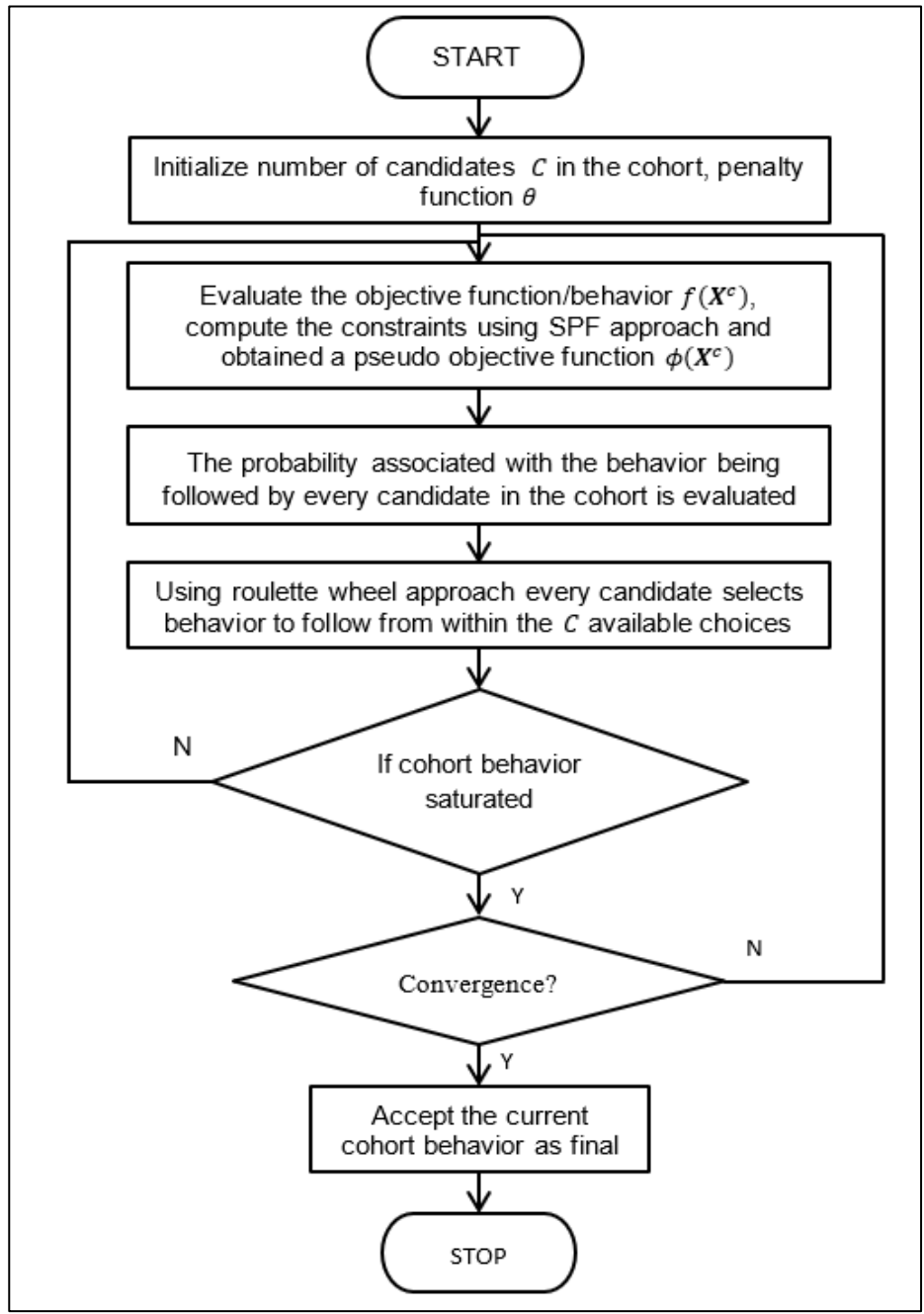

**Fig. 3 Cohort Intelligence basic working mechanism [1]**

The layout of this paper is section 2 discusses CI, section 3 its biblio-analysis and section 4 network analysis, followed by conclusion in section 5 and references used in the last section.

## 2. Cohort Intelligence (CI)

Kulkarni et al. in 2013 proposed this socio driven algorithm [1]. Learning based on social tactics through candidates involved in the cohort and competitive behavior among them is the key aspect here [16-18]. Behavior is exhibited by these candidates which sometimes lead to improvisation to their own behavior or behavior of the cohort. Specific qualities in a particular behavior makes candidate to adopt the same [20]. This in a way results in a continuous learning curve and overall cohort's learning will get evolved. When learning of all the involved candidates reaches to a peak then the cohort is said to be saturated or converged. In a way CI revolves around candidates, their continuous learning, exchange of the learned qualities, attainment of learned quantities to their peak [19-23].

Table 1 Detailed illustration of CI version along with its applications [1-35]

| Sr. No. | Evolved Version of CI | Year | Tested Applications/Problems |
|---|---|---|---|
| 1 | The basic version of CI algorithm | 2013 | Griewank, Rosenbrock, Ackley, Sphere are unconstrained test functions, and they are successfully tested. |
| | | 2017 | Parameter Optimization of PID Controller |
| | | 2018 | Heat exchanger with shell tube get optimized |
| | | 2019 | Cooperative multi-agent self-organizing robotic system is designed |
| 2 | Modified version of CI using K-Means (K-MCI) | 2014 | Data sets from UCI Machine Learning Repository |
| 3 | CI using probability based constrained handling approach | 2016 | Knapsack problems (0-1), real-world application such as inventory management for healthcare and, mix problem for a sea-cargo-mix problem and selection of a cross-border-shipper problems solved |
| 4 | CI along with AHP and MAHP approaches | 2016 and 2017 | Ice-cream suggestion to a diabetic patient |

| Sr. No. | Evolved Version of CI | Year | Tested Applications/Problems |
|---|---|---|---|
| 5 | Constrained version of CI with dynamic and static penalty approach | 2017 and 2019 | Test problems and mechanical design engineering problems |
| 6 | Variations of CI | 2018 | Mesh smoothing of complex geometry |
|  |  | 2019 | Designing fractional PID controller |
| 7 | CI with Cognitive Computing | 2018 | Improve the JPEG picture quality and steganography. |
| 8 | Multi-CI algorithm | 2018 | Tested on 75 test constrained problems |
|  |  | 2020 | manufacturing engineering problems |
| 9 | Pareto-dominance based Multi-objective Cohort Intelligence algorithm | 2020 | Test problems |
| 10 | Self-adaptive-cohort intelligence & Opposition-based learning | 2020 | Patients with no-show-data of a primary-care clinic |
| 11 | Modified-multi-cohort intelligence algorithm with panoptic-learning for unconstrained problems | 2020 | Unconstrained test problems |

Initially, Griewank, Rosenbrock, Sphere, and Ackley functions were tested using CI which were benchmark unconstrained test functions [1]. Then, 0-1 Knapsack and NP-hard problems with 4-75 number of varying items were tested for problems having constraints [2]. Same type of smaller size knapsack 0-1 problems was solved using CI algorithm [3]. A problem specific probability-based constraint handling technique handles the constraints involved in this problem. Integer-based programming-solutions competent results were yielded by the algorithm. In actual combinatorial problems from logistics and healthcare domain can also be solved by this way. The

Cross Border Supply-Chain domain problems which are of large size, complex was also tried to be solved by using this algorithm [4], several other benchmark problems [6], and Travelling Sales Problem (TSP) were also solved by using CI. The algorithm outperformed in comparison to the integer-programming and other type of in-particular heuristics mechanisms.

Modified-CI developed by Krishnasamy et al. using mutative approach acted as a utility for expanding the available sample-space through diverseness and in addition to this same used to avoid converging behavior of premature type [7]. The modified-CI was served better in relation to basic CI and showed utility towards solving of different types of clustering scenarios. Additionally, hybridized CI was designed by incorporating with K-Means' algorithm resulting into the superior performance. Gaikwad et al. proposed an Analytical Hierarchy Process (AHP) and Modified AHP which was combo of GA & CI and showed effectualness towards diabetic patients ice-cream recommendation [8,9].

The dynamic and static penalty function approaches were incorporated in CI-DPF and CI-SPF for solving manufacturing engineering problems and for solving several test problems too [10,11]. Truss structure and mechanical engineering domain [11-14] are for complex type and solved using CI-SPF. CI was also used in order to optimize process parameter of water abrasive jet machining [15]. Furthermore, seven variations to a CI algorithm were introduced by Patankar and Kulkarni [16]. Candidates' choice to select other candidate from which to learn certain characteristics are the variations which were incorporated by them. Roulette wheel selection, follow better, follow best, follow itself, follow worst, random selection, alienation and follow median are the variations. Multimodal unconstrained problems and several unimodal problems were tested using it. These variations are then further utilized to mesh smoothing by minimizing the condition number [17]. The base version of CI algorithm was utilized to optimize the PID parameter in DC motor and to design the PID fractional controller [18,19]. Then variations in CI algorithm were also design of fractional PID controller [20]. In addition to these, through steganography for the security of private messages CI was also applied [21,22]. The cryptography algorithm based on CI was produce a strong cipher text to encrypt the secret messages [23]. Further, CI algorithm was implemented using Cognitive Computing (CICC) for JPEG image steganography for small quantization tables [24]. Then it was again modified for gray scale image having high-capacity quantization tables using reversible data hiding scheme [25,26].

The shell-and-tube heat-exchanger design issues were also resolved for reduction in cost and yielded relatively better results as compared to other algorithms of the considered category [27]. The CI was hybridized with Adaptive Range Genetic Algorithm (ARGA) and tested on fifty benchmark test problems and shell-tube heat-exchanger design issues [28]. Furthermore, a Multi-CI algorithm was proposed based on the interaction among different cohorts and it was validated on 75 unconstrained test problems [29] and advanced manufacturing engineering problems [30]. Further, the Multi-CI was modified using panoptic learning and tested using unconstrained benchmark problems.

Moreover, an exponential probability based multi-agent CI algorithm was proposed to run the swarm robotic application based on No Obstacle Case (NOC), Rectangular Obstacle Case (ROC), Multiple Rectangular Obstacles Case (MROC) and Cluttered Polygonal Obstacles Case (CPOC) [31]. The CI algorithm was also used to minimize the clearance variation of radial selective assemblies of hole and shaft [32]. A CI based strategy was introduced to deal with the applications of load power supply in the reconstruction of a traditional ship power distribution information network (SPDN) for maximum power recovery [33]. A new approach was proposed for the healthcare system based on three variants of Opposition-based Self-Adaptive Cohort

Intelligence (OSACI) algorithm such as OSACI-Init, OSACI-Update and OSACI-Init_Update [34]. Finally, a novel pareto dominance based Multi-objective CI (MOCI) algorithm was proposed [35] and tested on several benchmark problems and real-world applications.

The outline of this paper is section 2 covers preliminary information, followed by bibliometric analysis in section 3, conclusion in section 4 and references used to formulate this paper be at the end.

## 3. Cohort Intelligence Biblio-Analysis

As discussed in the introduction and Cohort Intelligence sections, it is an emerging algorithm, So, it is necessary to get an idea about the details about it from the point of view of Scopus and Web of Science.

The first important detail is the query fired to Scopus and Web of Science databases. The query is details are as follows.

```
( "Cohort Intelligence" AND "artificisal intelligence" OR "sociso-inspired" OR "K-Means" OR "Knapsack" OR "Healthcare and logistics" OR "Roulette Wheel" OR "Discrete and Mixed Variable" OR "Truss Structure" OR "Design Engineering" OR "Cross Border Supply Chain" OR "Probability" OR "Static and Dynamic Penalty" OR "Feasibility Based Rule" OR "Self-adaptive penalty function" OR "colliding bodies" OR "PID-controller" OR "Advanced Manufacturing" OR "variations off ci" OR "multi cis" OR "Multi Objective" OR "ci-gas" OR "cis-ahp" OR "Shell tube" OR "JPEG" OR "Finite Element" )
```

**Fig. 4 Query given as an input to Scopus and WoS database**

26 documents are obtained from Scopus and 18 from Web of Science.

### 3.1 Yearly Document Count

Yearly document count w.r.t. Scopus and WoS are shown in fig. 5 and 6. There is a growth in CI publications from 2015 to 2019. Since, 2014 research about CI is in inception. In 2018, there are maximum number of WoS publications. In WoS also CI related research started from 2014. In a way fig. 4 and 5 somewhat similar patterns for year wise document count. It is obvious that the count in these two databases may vary as it is not the case that those publications that are in Scopus need not to be indexed in WoS database.

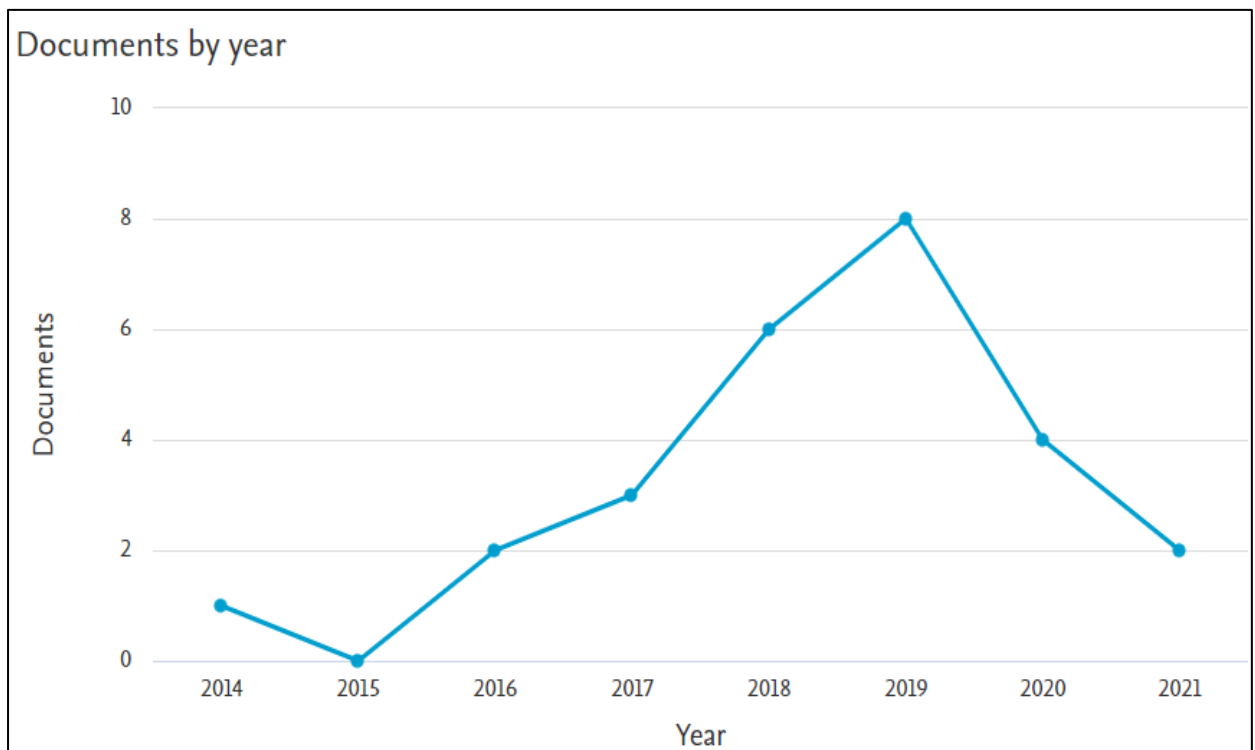

**Fig. 5 Yearly Publication Count (Scopus_DB Perspective)**

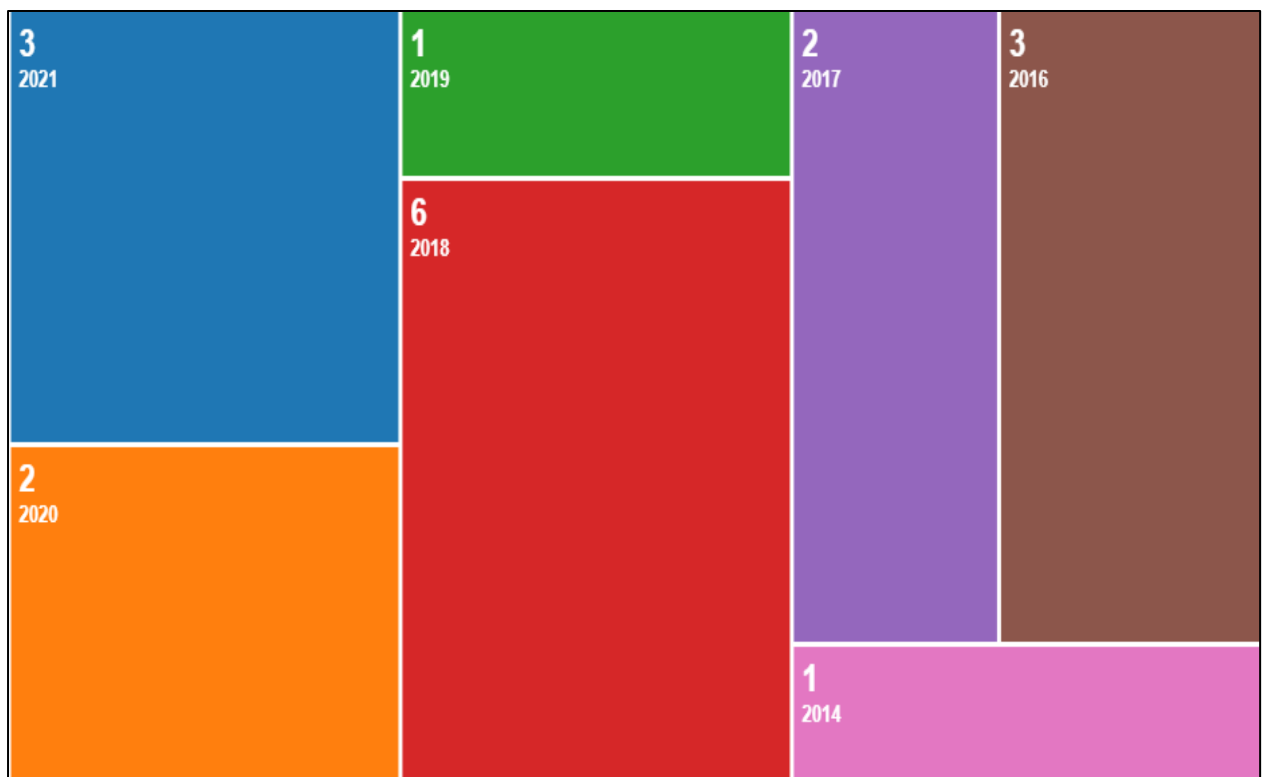

**Fig. 6 Yearly Publication Count (WoS_DB Perspective)**

## 3.2 Type wise Document Count

From fig. 7 and 8, contribution of articles is more. There are early assess type documents also. No Contribution in terms of review papers. So, there is a scope for review articles.

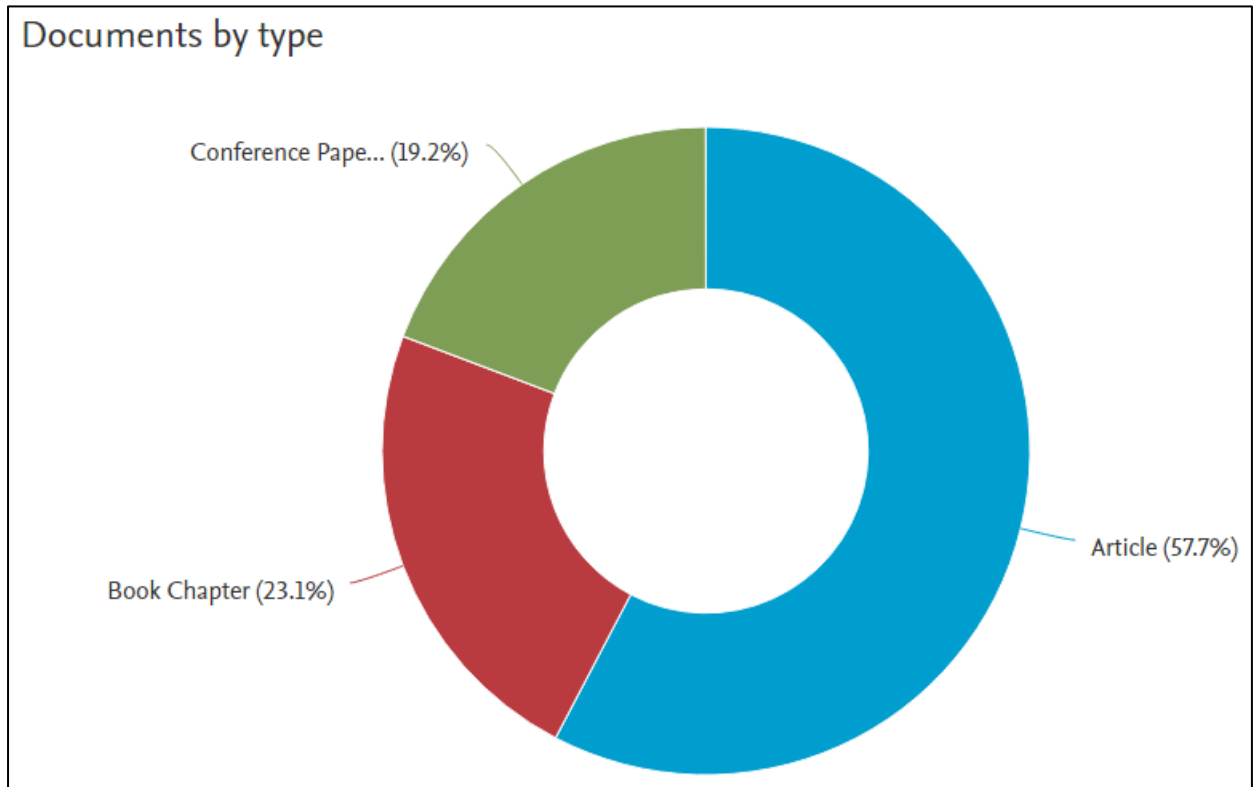

**Fig. 7 Type wise Document Count (Scopus_DB Perspective)**

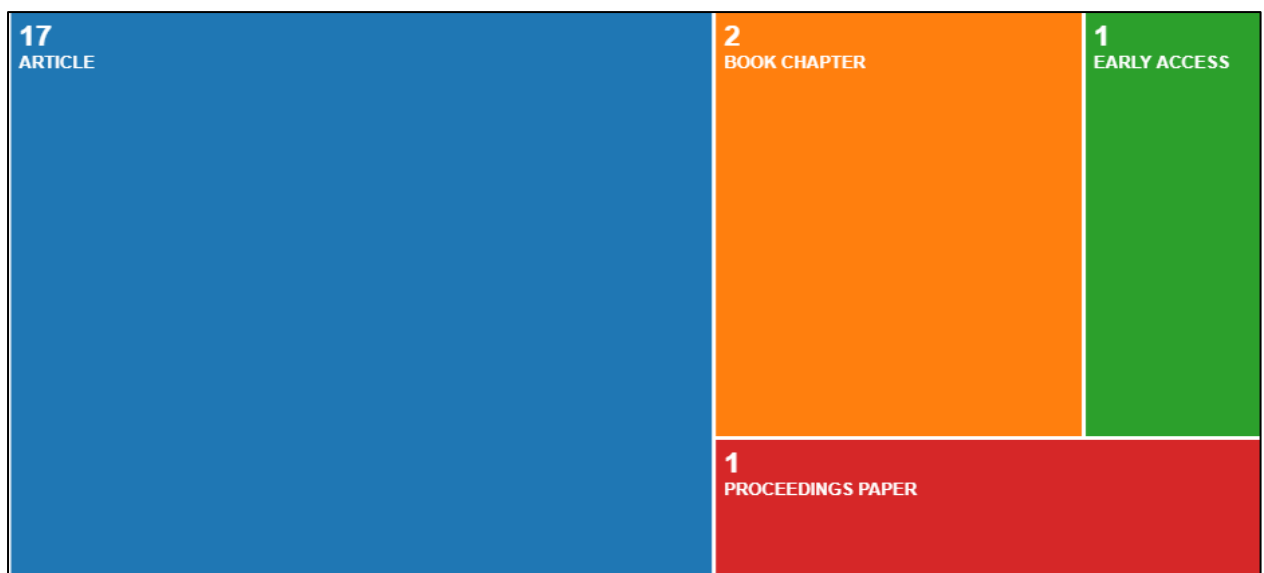

**Fig. 8 Type wise Document Count (WoS_DB Perspective)**

## 3.3 Subject area wise Documents Segregation

From fig. 9 and 10, contribution of computer science area is more. Other involved subject areas. Other involved subject areas are physics and astronomy, neuroscience, social science, mathematics and decision sciences, engineering etc. It shows diverse applicability of CI algorithm.

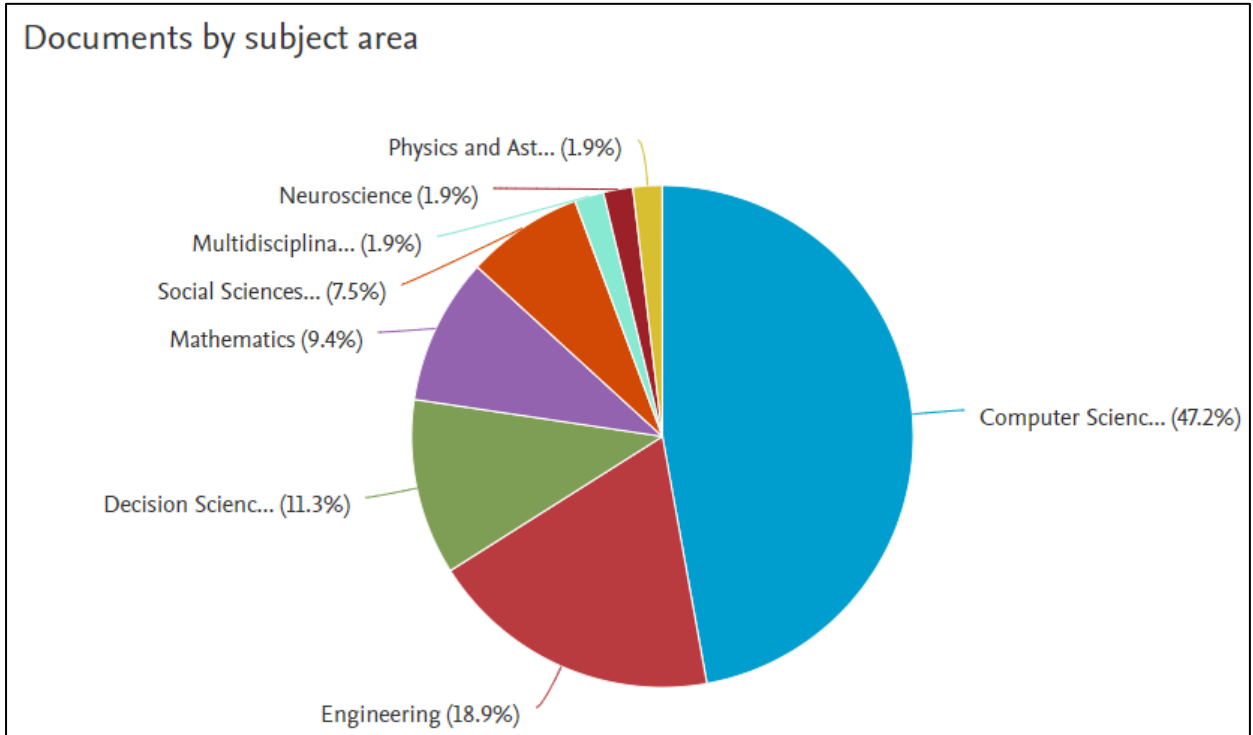

**Fig. 9 Classification of Documents as per Subject Area (Scopus_DB Perspective)**

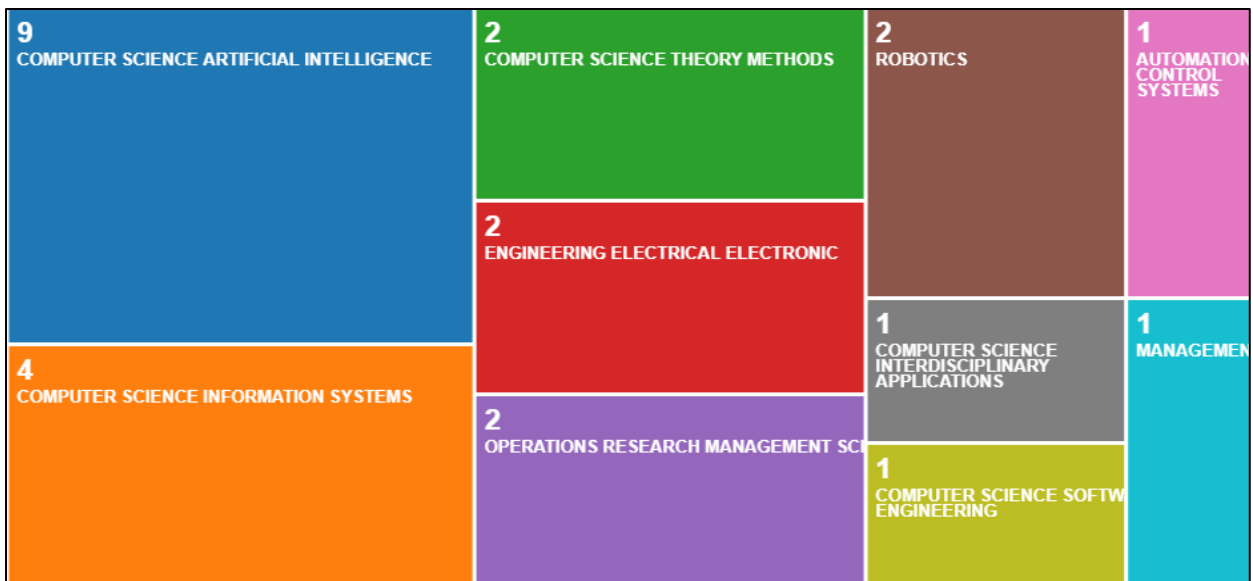

**Fig. 10 Classification of Documents as per Subject Area (WoS_DB Perspective)**

## 3.4 Affiliation specific Documents

Top two affiliations are Symbiosis International (Deemed University) and University of Windsor for CI related research.

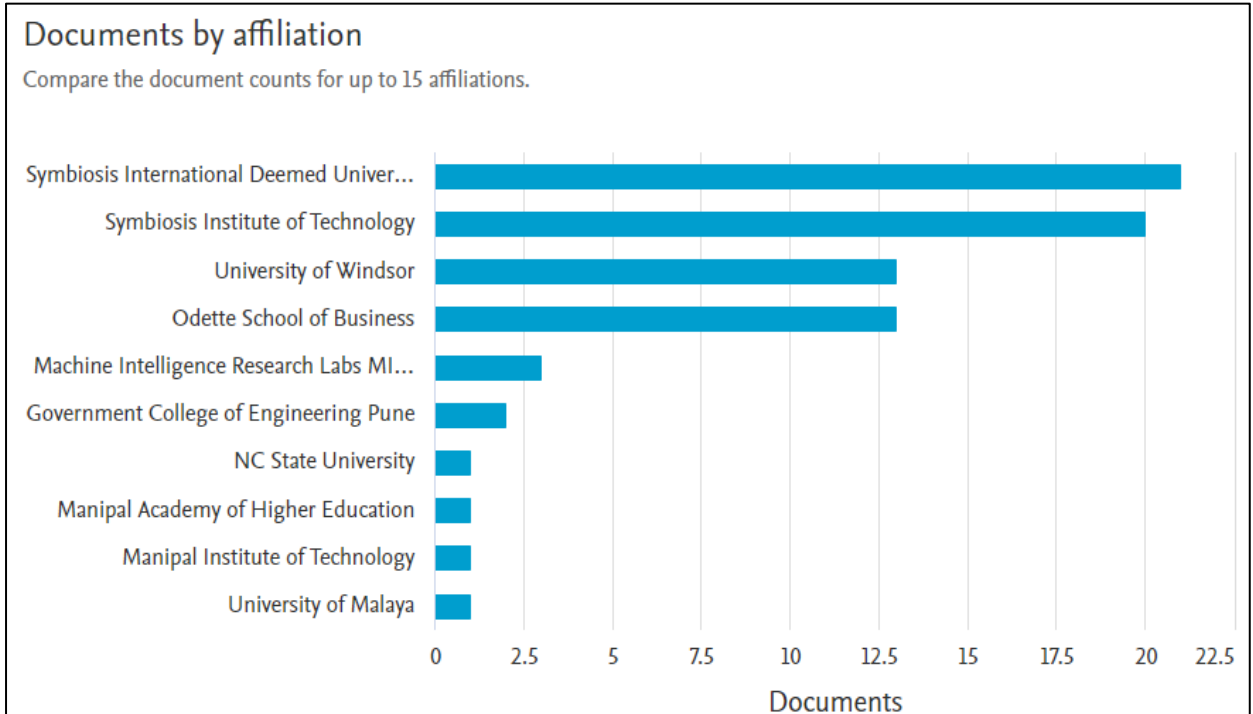

Fig. 11 Affiliation specific Documents (Scopus_DB Perspective)

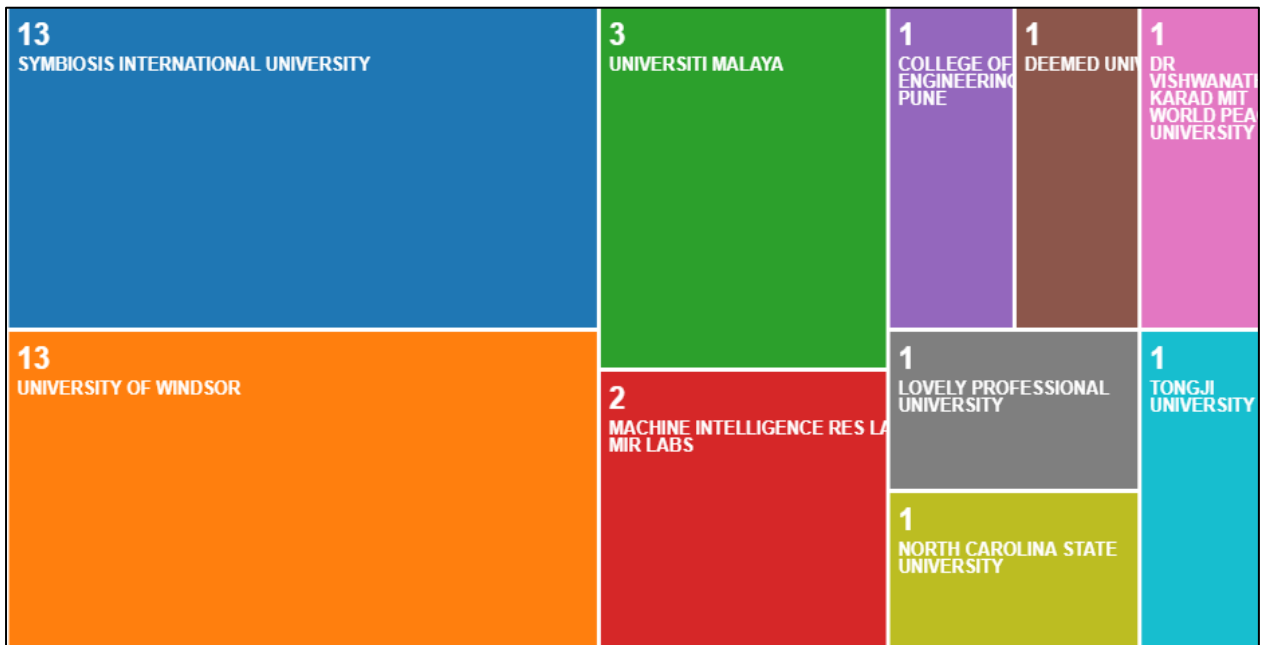

Fig. 12 Affiliation specific Documents (WoS_DB Perspective)

## 3.5 Author specific Document Count

Fig. 13 and 14 shows that Kulkarni A. J. and Sarmah D. K. are the top contributing authors for CI related research.

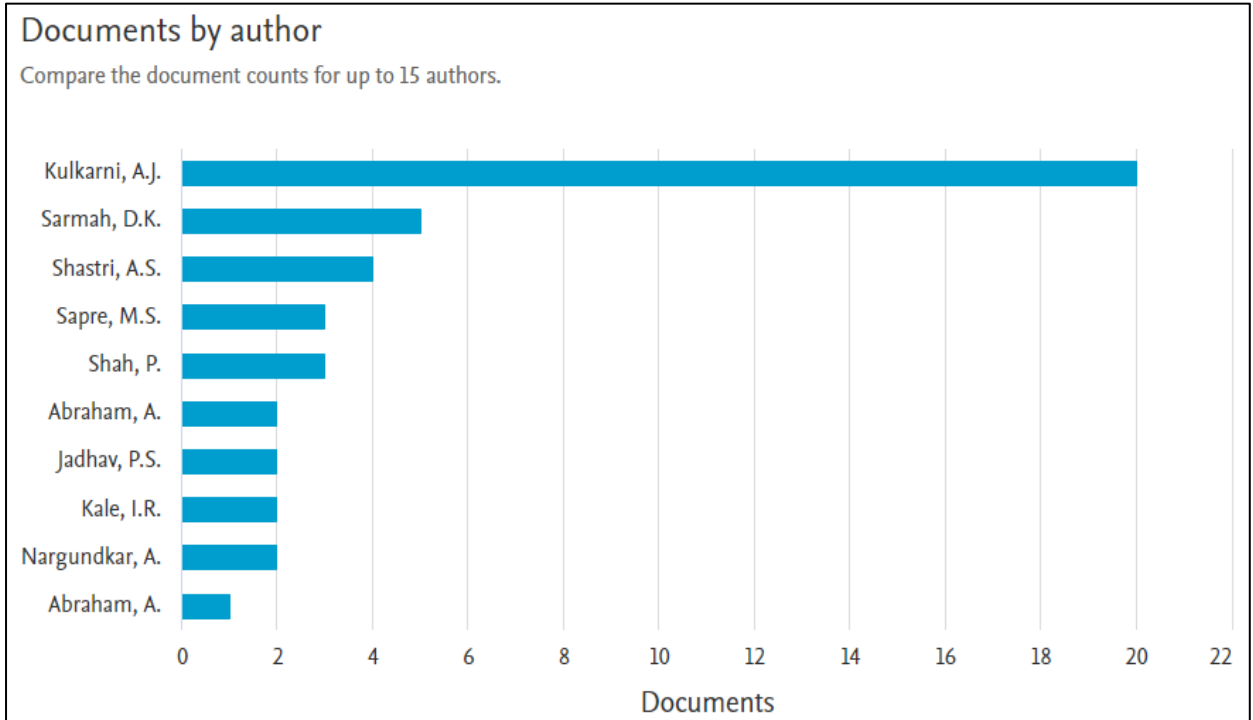

Fig. 13 Author specific Document Count (Scopus_DB Perspective)

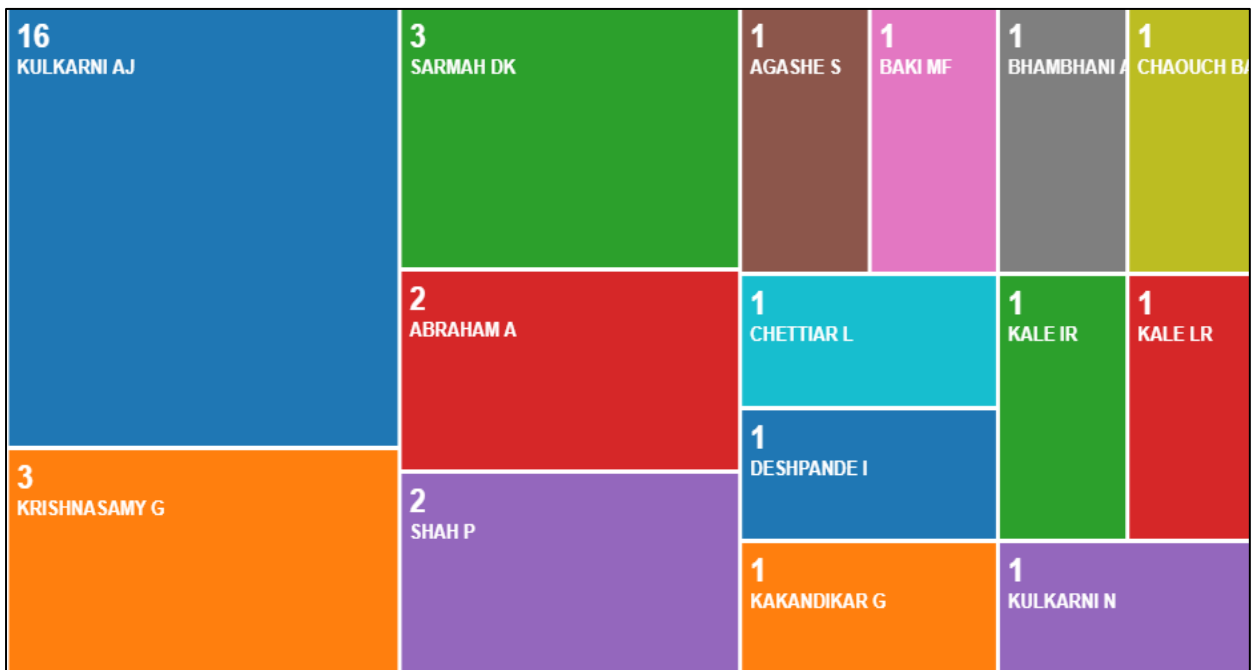

Fig. 14 Author specific Document Count (WoS_DB Perspective)

## 3.6 Territory specific Document Count

Fig. 15 and 16 shows that CI related documents are more in number from India and Canada. This in a way confirms that more contributing authors are from India and Canada affiliations.

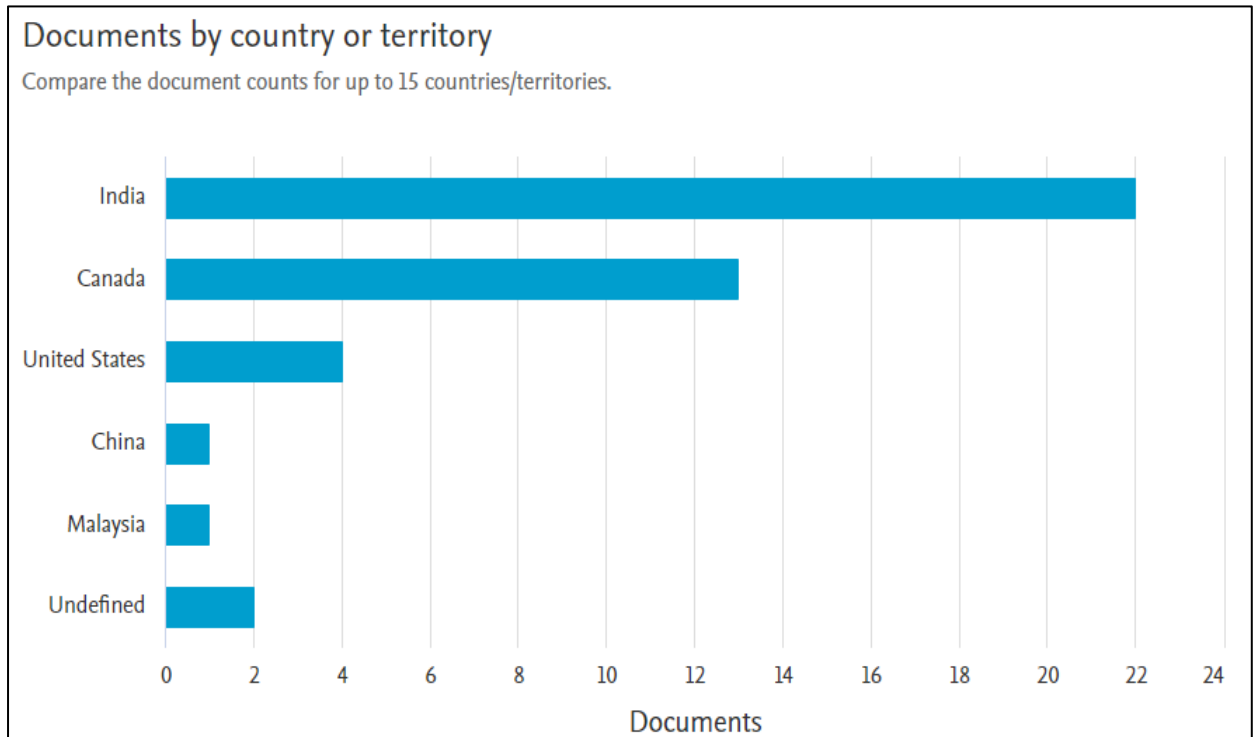

**Fig. 15 Territory specific Document Count (Scopus_DB Perspective)**

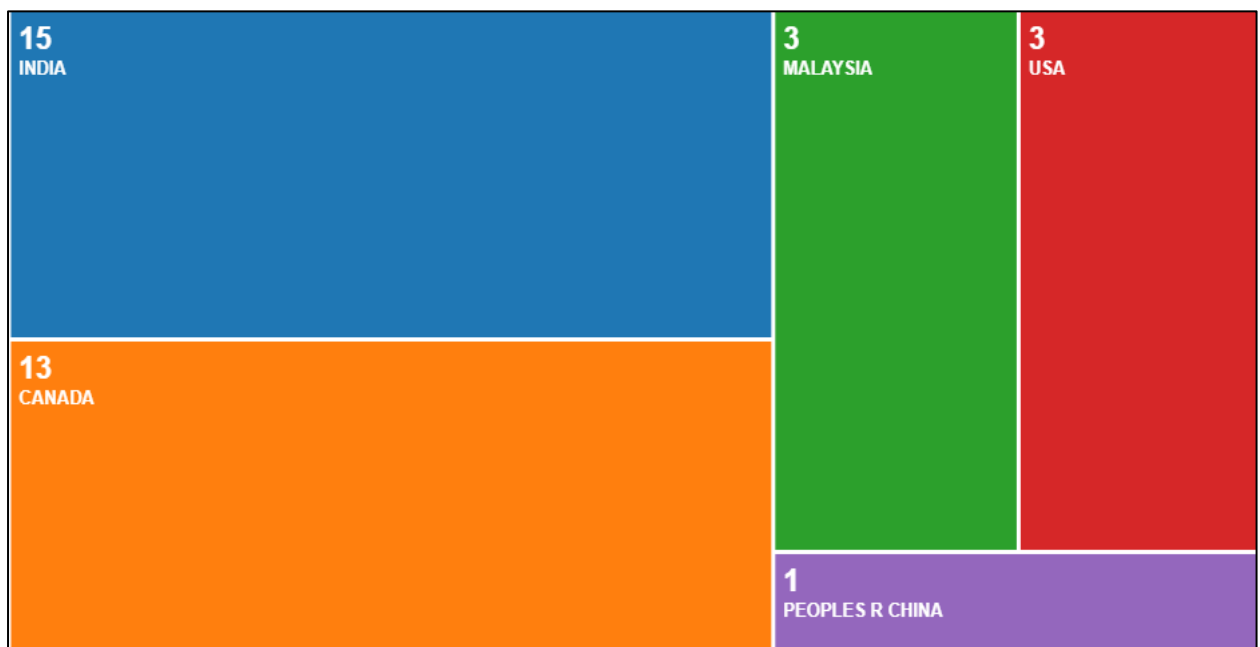

**Fig. 16 Territory specific Document Count (WoS_DB Perspective)**

**3.7 Document source specific Count**

Fig. 17 and 18 shows that top document sources are ISRL (Intelligent Systems Reference Library), Information Sciences, Studies in Computational Intelligence etc.

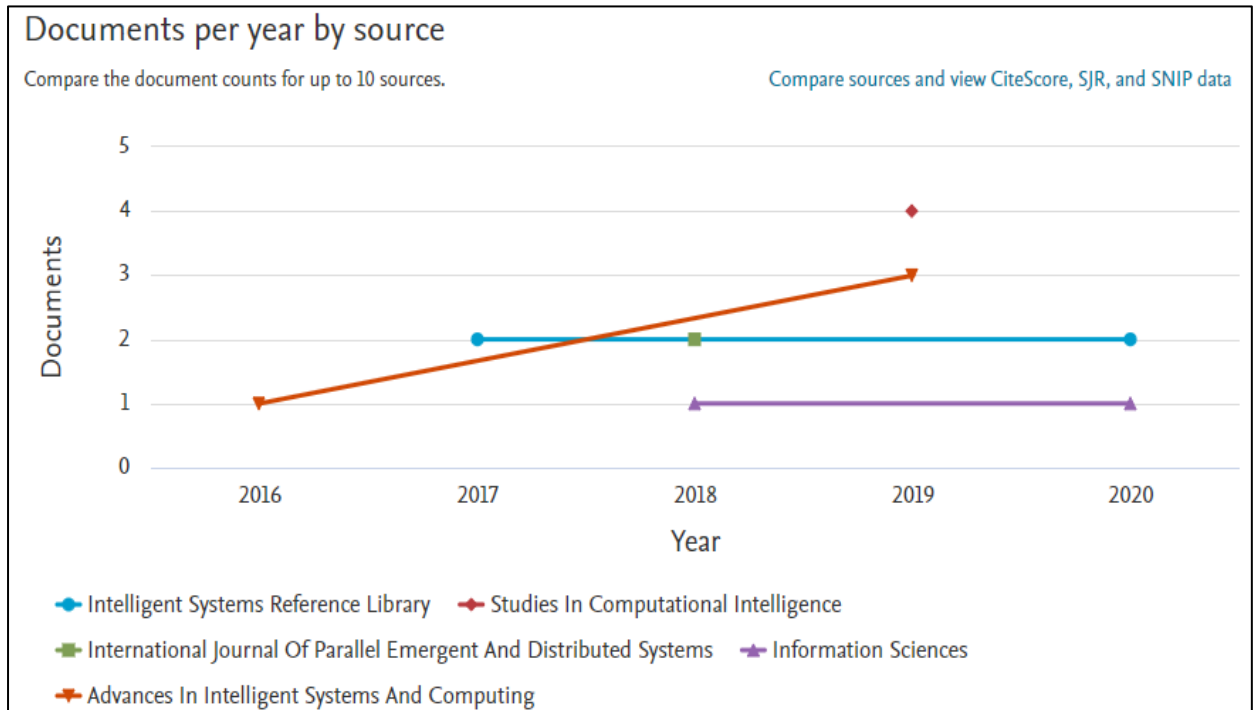

Fig. 17 Document source specific Count (Scopus_DB)

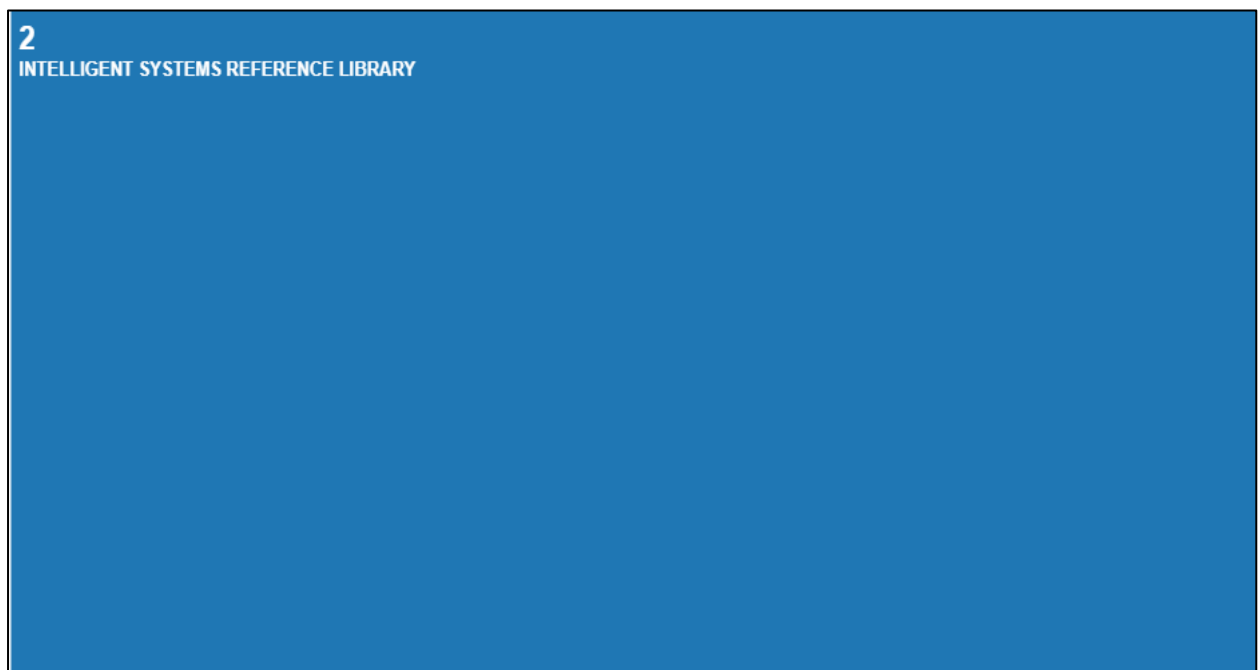

Fig. 18 Document source specific Count (WoS_DB)

## 3.8 Funding Sponsor specific Documents

Fig. 18 shows funding sponsor specific documents. Two funding sponsors viz. Ministry of Higher Education, Malaysia and National Natural Science Foundation of China funded one research documents each.

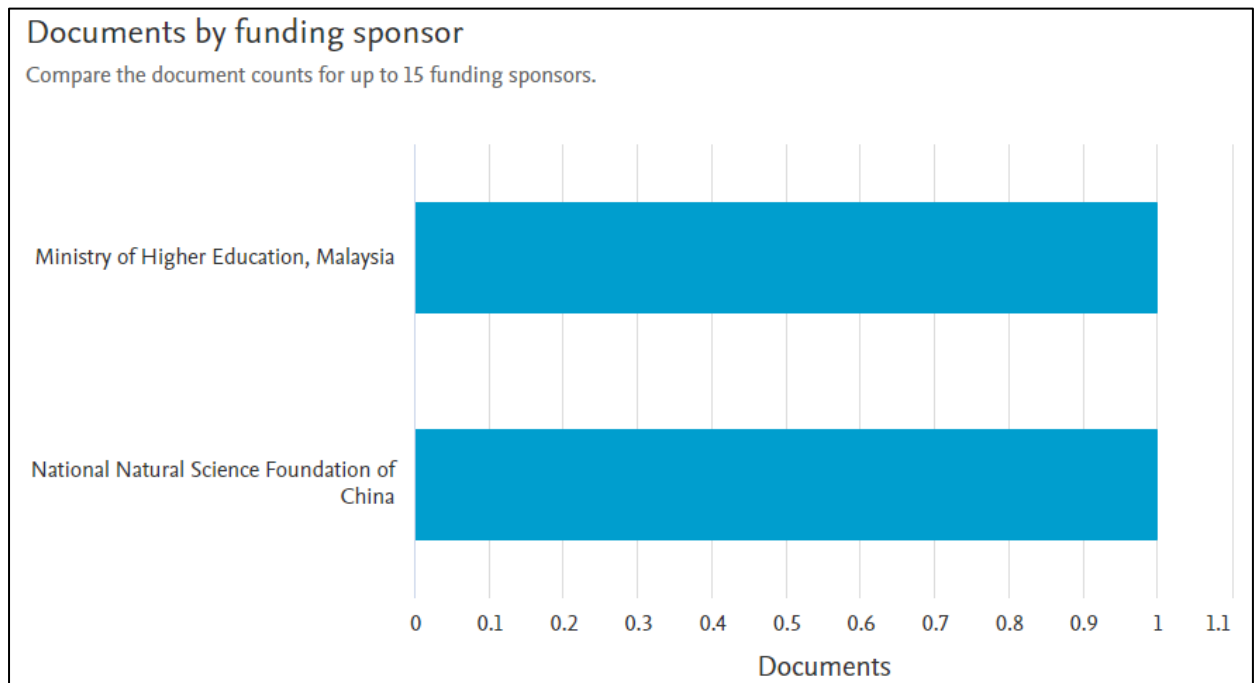

Fig. 18 Funding Sponsor specific Documents (Scopus_DB)

## 4 Network Diagrams for Cohort Intelligence related Research

### 4.1 Authors-Keywords Coappearance

Cohort Intelligence is the main keywords and other coappearing keywords are constrained optimization, combinatorial optimization, 0-1 Knapsack problem, condition number, genetic algorithm, hexahedral mesh. The authors showing coappearance for these keywords are along with Kulkarni A. J. are Kale I. R., Sarmah D. K., Shastri A. S., Shah P., Sapre M. S., Nargundkar A. etc. These authors belong to the same affiliation that is Symbiosis International (Deemed University). It is also observed that Kulkarni A. J. is the PhD supervisor of these mentioned authors.

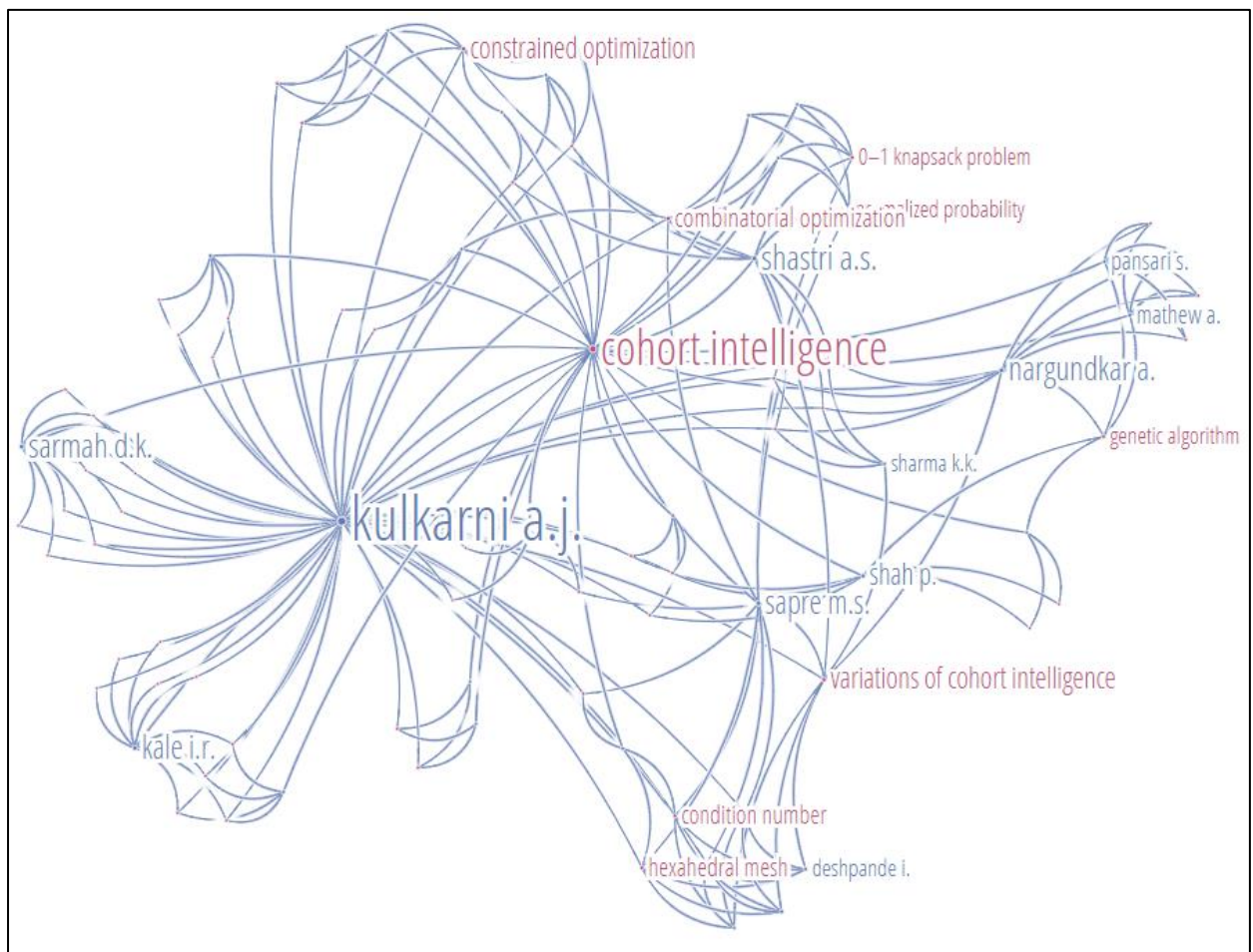

**Fig. 19 Authors-Keywords Coappearance (Scopus_DB)**

## 4.2 Authors_Source_Titles Coappearance

It is important to get an idea about the source titles related to CI related research. The source titles shown in fig. 20 are studies in computational intelligence, neural computing and applications, advances in intelligent systems and computing, expert systems with application, intelligent systems reference library, evolutionary intelligence, International Journal of Parallel, Emergent and Distributed Systems, Information Sciences etc. The coappearing authors for these journal titles are Kulkarni A. J. and Sarmah D. K. Kulkarni A. J. is leading author has he had developed CI algorithms and further research continued under his supervision. Kulkarni A. J. has his research lab also called OAT Research Lab and he did his PhD from NTU Singapore.

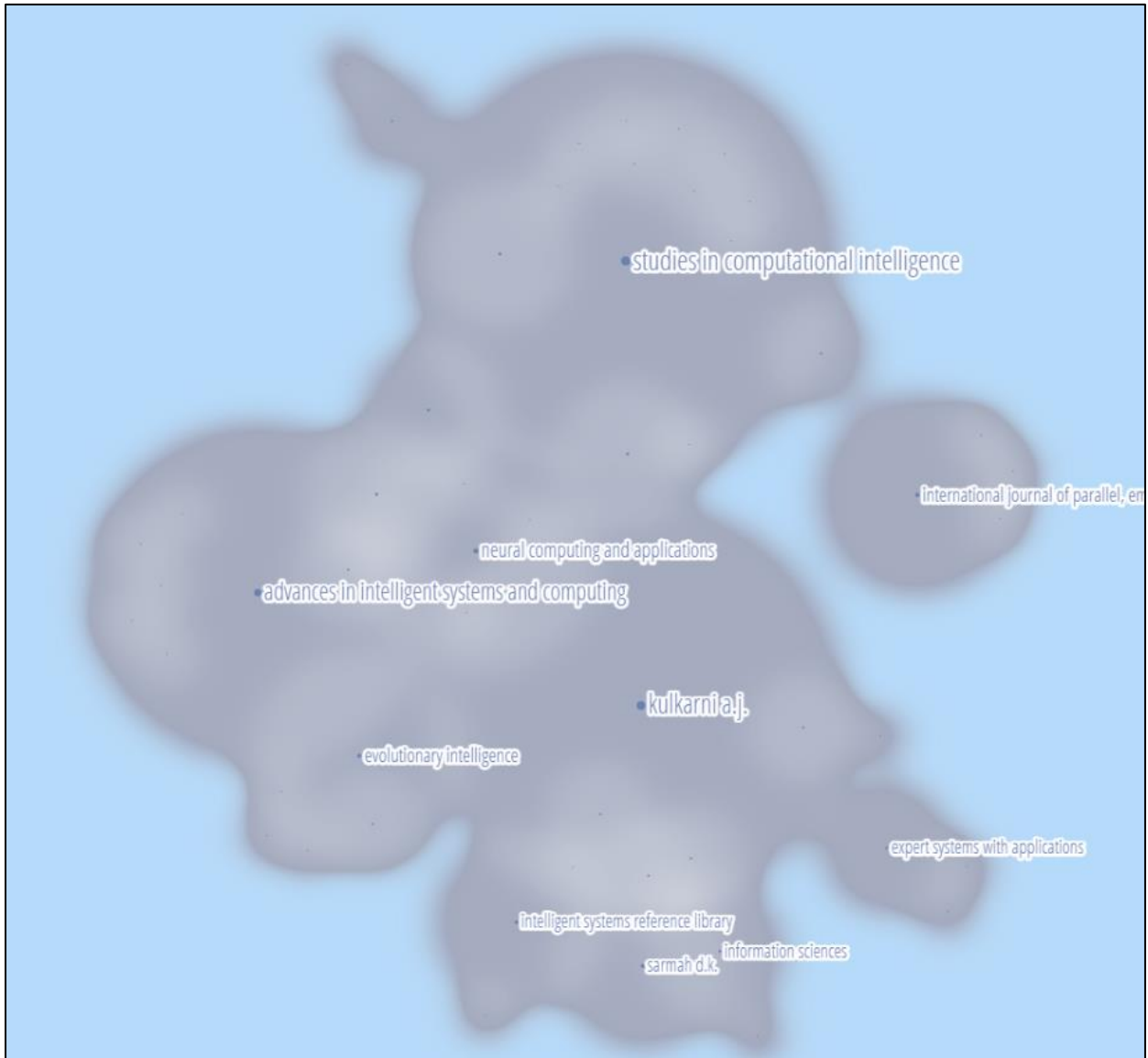

Fig. 19 Authors_Source_Titles Coappearance (Scopus_DB)

**4.3 Source_Titles_Keywords Coappearance**

Fig. 20 shows additional important keywords and source titles. Additional important keywords are dynamic penalty function approach, response surface methodology, normalized probability, engineering design optimization, optimization-based smoothing etc. Additional related source titles are Arabian journal for science and engineering, international journal of machine learning and cybernetics, neural computing, and applications etc. Through this network diagrams other important facets related to CI research are obtained.

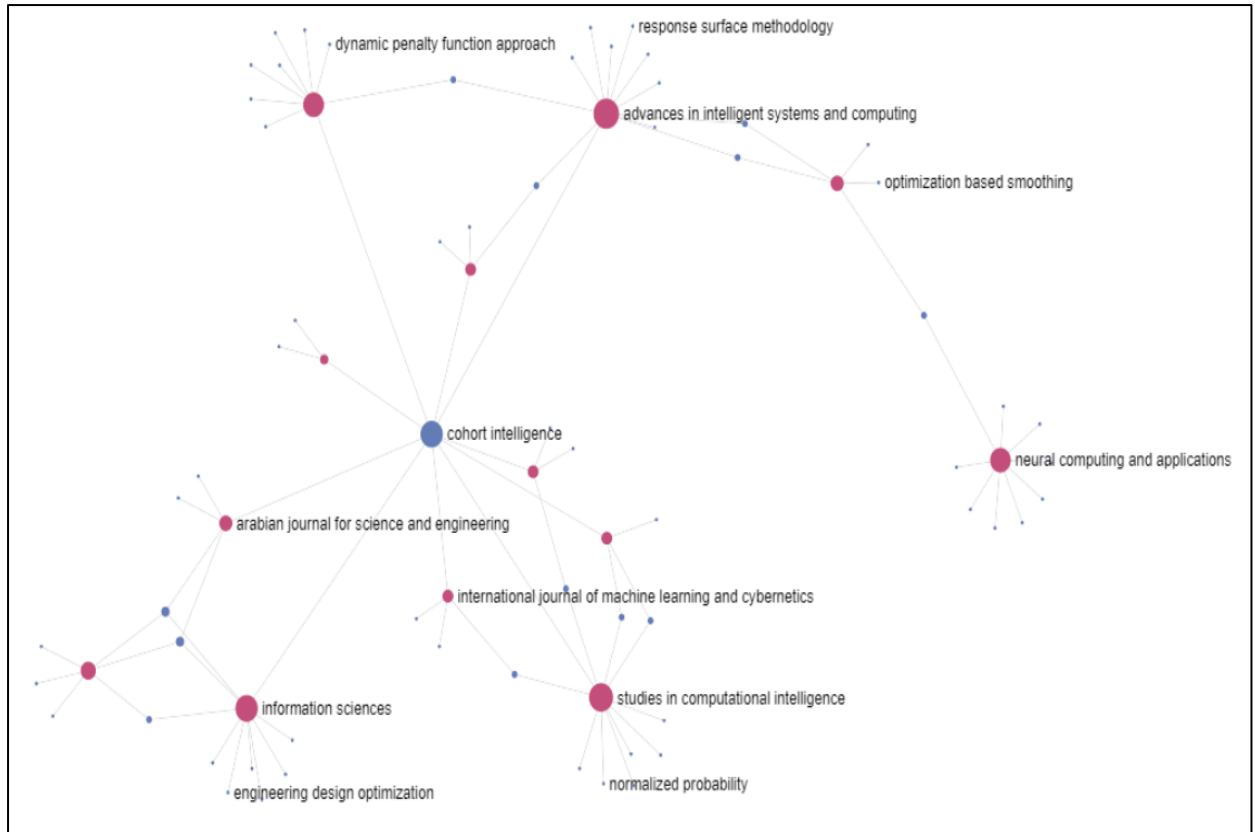

**Fig. 20 Source_Titles_Keywords Coappearance (Scopus_DB)**

**4.4 Authors Co-publications**

Authors having linkages through their publications shows additional authors involved in CI related research. Shabir H., Mehra S., Sapre M. S., Thorat E. V., Thaker R., Patel H., Piprikar B., Chettiar I., Paramesran R. etc. So, in a way these are the additional authors who worked with Kulkarni A. J. This type of analysis is also important as in previous sections we got an idea about top authors from Scopus and WoS databases point of views and here authors doing networked research are mentioned.

In a way network diagrams gives more insights than those that are given by the popular publication databases. It is important to consider these additional insights as for a new researcher or for an already doing research in the said area author or the researcher such type of details act as a catalyst.

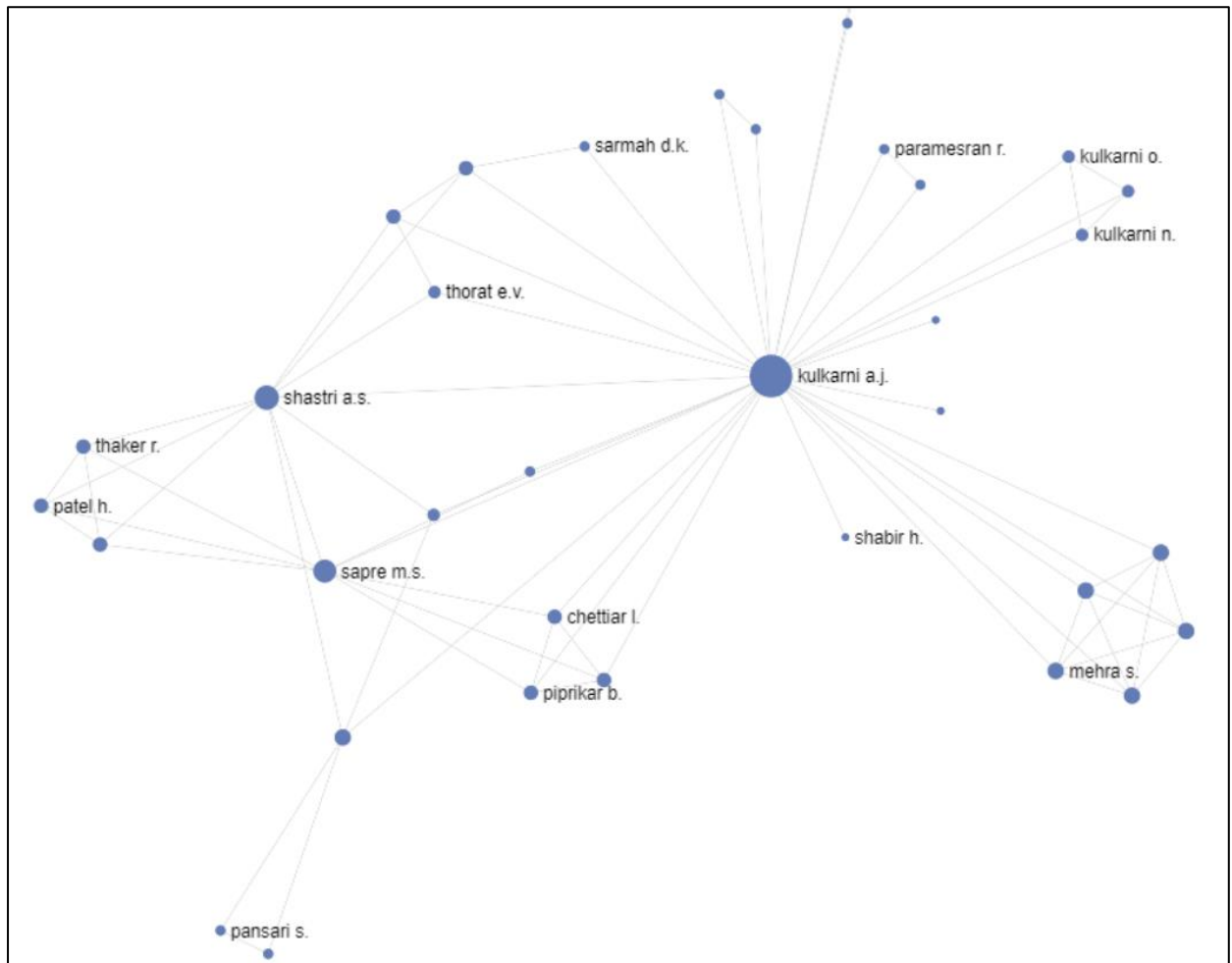

**Fig. 21 Authors Co-publication (Scopus_DB)**

## 4.5 Citations linkage among publications

Fig. 22 is self-explanatory. It shows publications related to CI. Also, this figure tries to capture all research articles related to CI.

## 4.6 Keyword Coappearance

Keywords showing coappearances are abrasive water jet machining, metaheuristics algorithms, self-supervised learning, constrained optimization, constrained test problems, fractional calculus,

clustering, pid, genetic algorithm, discrete cosine transform, steganography, multi random start local search, dynamic penalty function approach, discrete and mixed variables etc.

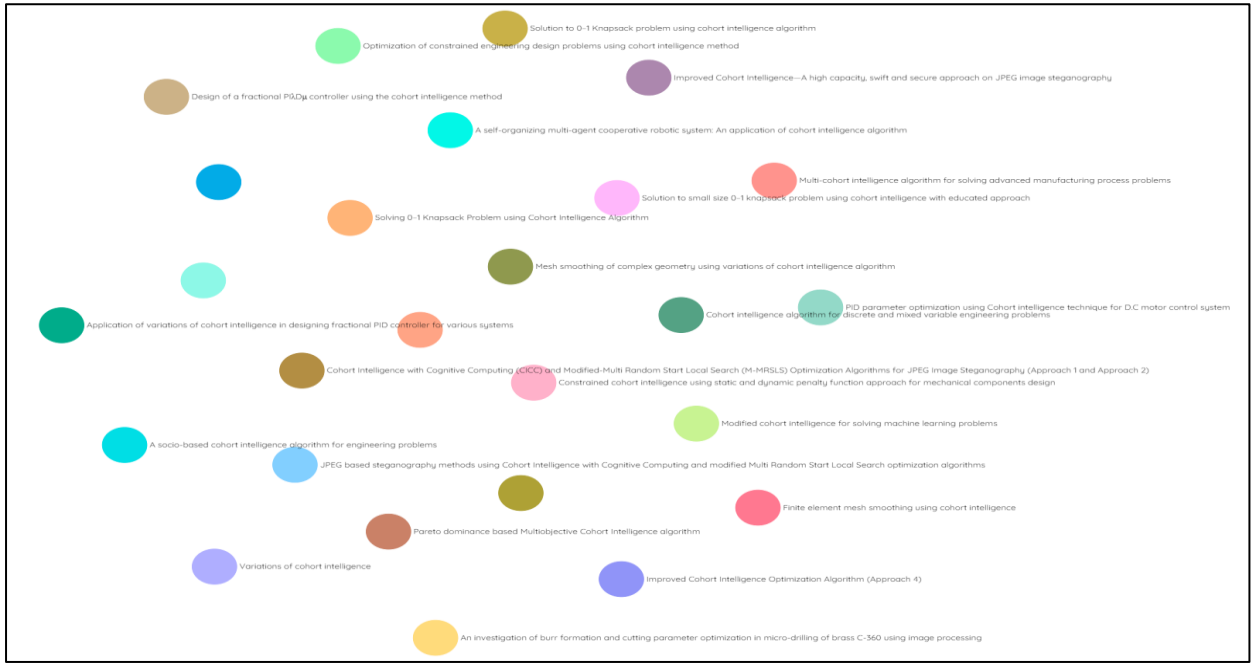

Fig. 22 Publications linked through Citations (Scopus_DB)

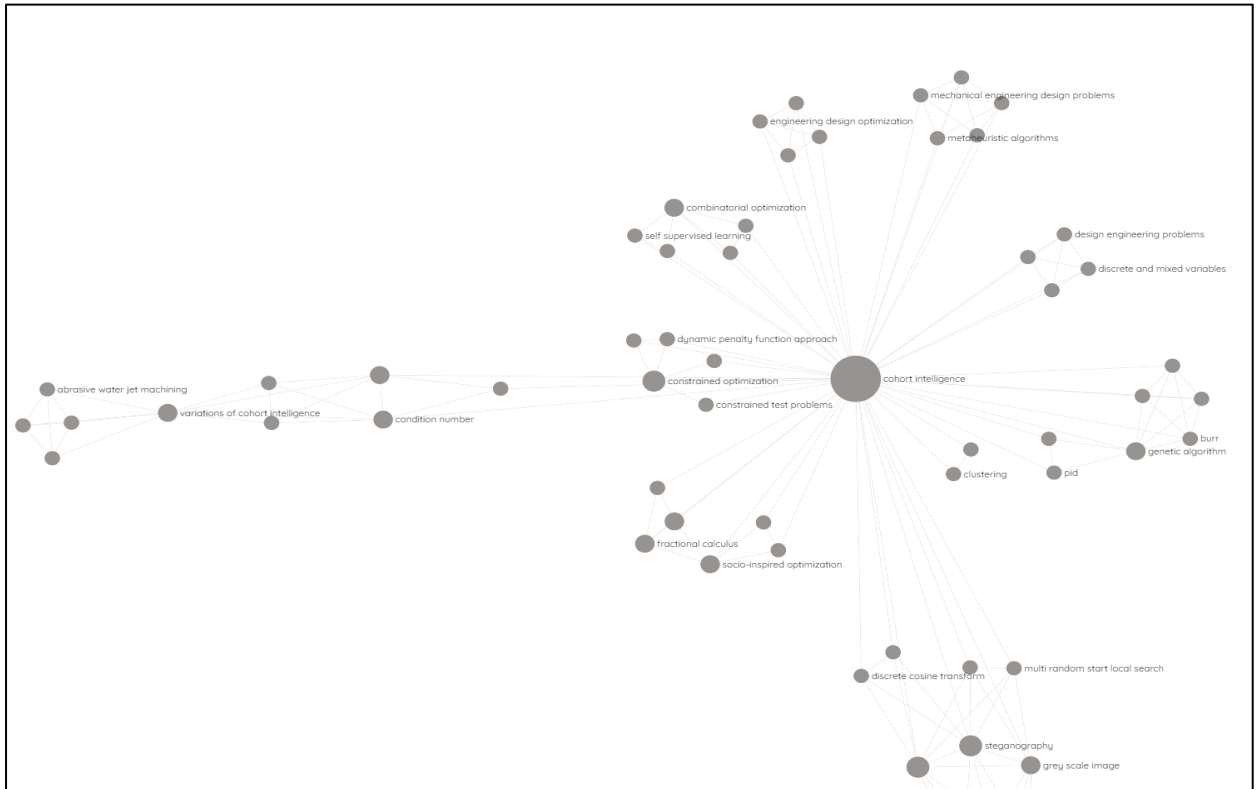

Fig. 23 Keywords_Coappearance (Scopus_DB)

## 4.7 Citations overview from Scopus_DB

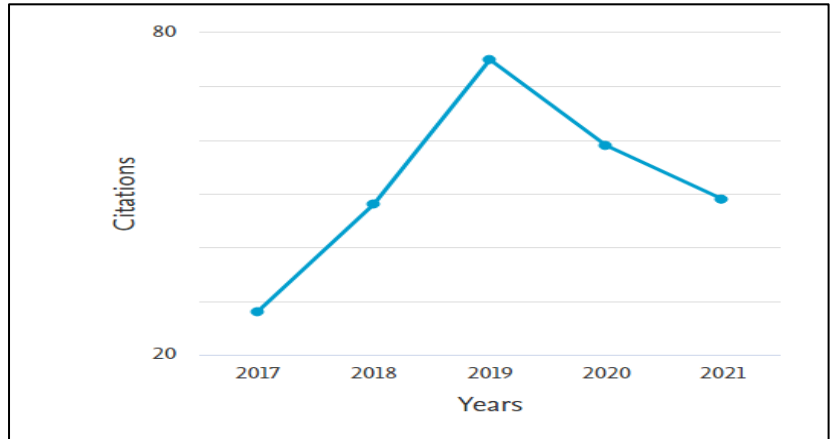

**Fig. 24 Overall Citation Overview**

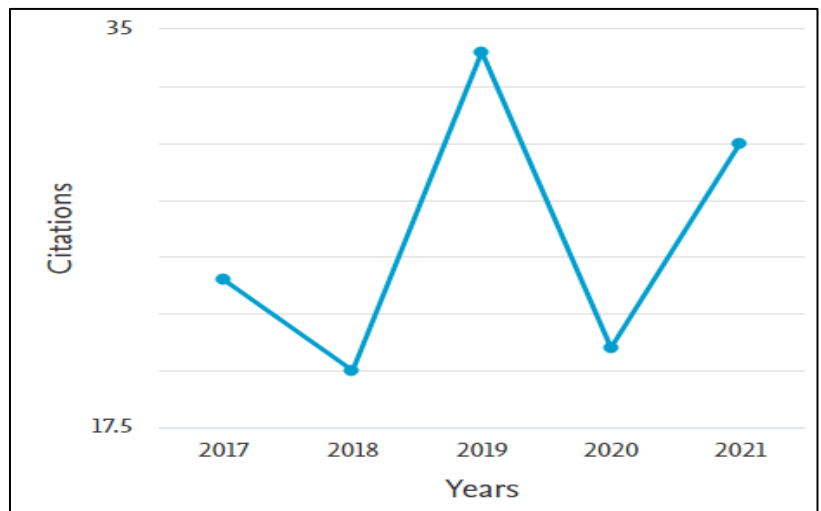

**Fig. 25 Citations overview after inclusion of self-citations**

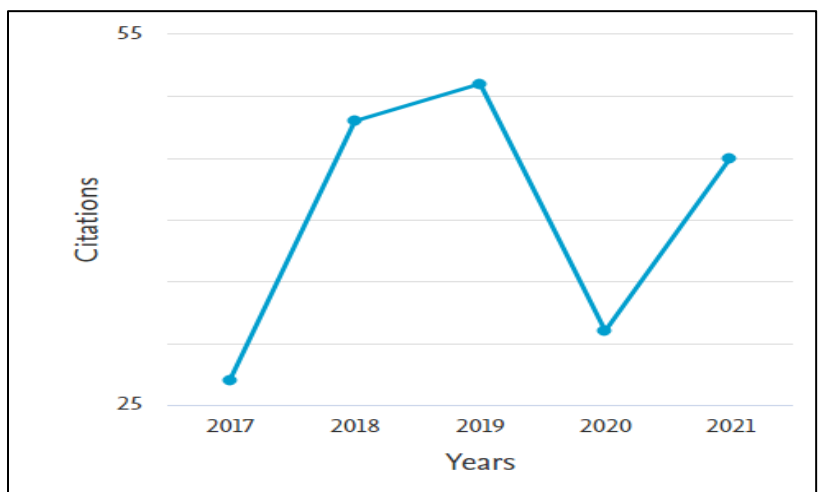

**Fig. 26 Citations overview after inclusion of citations from books**

| | Documents | Citations | <2017 | 2017 | 2018 | 2019 | 2020 | 2021 | Subtotal | >2021 | Total |
|---|---|---|---|---|---|---|---|---|---|---|---|
| | | Total | 20 | 27 | 48 | 51 | 31 | 45 | 202 | 0 | 222 |
| 1 | A hybrid approach for data clustering based on modified coho... | 2014 | 17 | 14 | 15 | 9 | 8 | 3 | 49 | | 66 |
| 2 | Solving 0–1 Knapsack Problem using Cohort Intelligence Algor... | 2016 | 3 | 12 | 14 | 12 | 2 | 7 | 47 | | 50 |
| 3 | Constrained cohort intelligence using static and dynamic pen... | 2018 | | | 6 | 4 | 1 | 2 | 13 | | 13 |
| 4 | Design of a fractional PI^λD^μ controlle... | 2018 | | | | 4 | 3 | 5 | 12 | | 12 |
| 5 | JPEG based steganography methods using Cohort Intelligence w... | 2018 | | | 4 | 3 | 3 | 2 | 12 | | 12 |
| 6 | Image Steganography Capacity Improvement Using Cohort Intell... | 2018 | | | 2 | 2 | 3 | 3 | 10 | | 10 |
| 7 | Variations of cohort intelligence | 2018 | | | 2 | 3 | 1 | 3 | 9 | | 9 |
| 8 | Improved Cohort Intelligence—A high capacity, swift and secu... | 2019 | | | | 4 | | 3 | 7 | | 7 |
| 9 | Cohort intelligence algorithm for discrete and mixed variabl... | 2018 | | | | 3 | 2 | 2 | 7 | | 7 |
| 10 | Solution to small size 0–1 knapsack problem using cohort int... | 2019 | | | | | 2 | 3 | 5 | | 5 |

**Fig. 27 Top 10 Citing Research Papers**

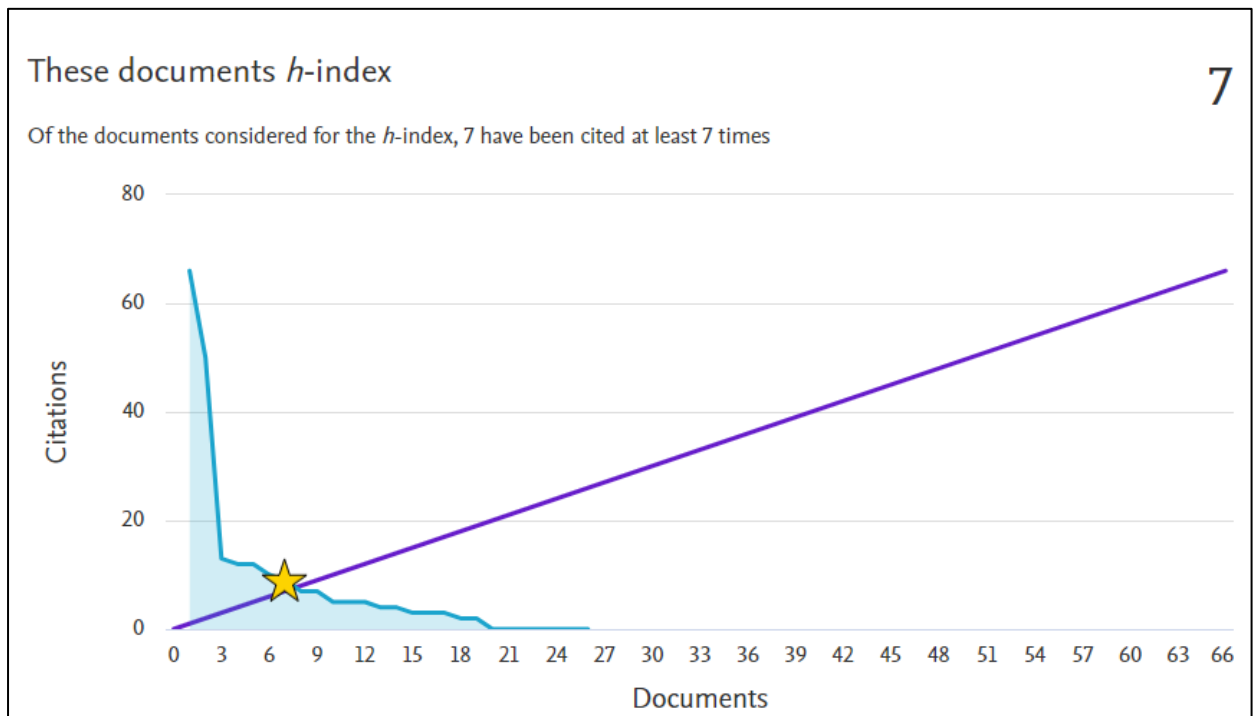

**Fig. 28 H-index of CI Research Papers**

### 4.8 Keywords Statistics

Fig. 29 and 30 shows keywords usage. Fig. 29 shows keywords which are frequently used in the considered research domain. On the other hand, fig. 30 shows less used keywords. Less used keywords in a way indicate that the more research by considering them need to carried out.

| | |
|---|---|
| Cohort Intelligence | (17) |
| Intelligence Algorithms | (6) |
| Particle Swarm Optimization (PSO) | (6) |
| Constrained Optimization | (5) |
| Genetic Algorithms | (5) |
| Optimization | (5) |
| Real-world Problem | (4) |
| JPEG Compression | (3) |
| Problem Solving | (3) |
| Simulated Annealing | (3) |

Fig. 29 Top 10 most Used Keywords

| | |
|---|---|
| Multi-objective Optimization | (1) |
| Multi-objective Problem | (1) |
| Multi-random Start Local Search | (1) |
| Multiobjective Optimization | (1) |
| Multiple Correlation | (1) |
| Multiple Performance Metrics | (1) |
| NP-hard | (1) |
| Nature Inspired Algorithms | (1) |
| Non-parametric Statistical Tests | (1) |
| Nonlinear Constrained Optimization | (1) |

Fig. 30 Top 10 Least Used Keywords

**4.9 Word Cloud for Top Cited Papers**

Fig. 31 and 32 shows word clouds for two most cited research papers. First paper has 75 and other has 59 citations. The main moto behind showing word cloud here is that to shows terminologies used in these two papers.

**Fig. 31 Word Cloud for "A hybrid approach for data clustering based on modified cohort intelligence and K-means"**

**Fig. 32 Word Cloud for Solving 0–1 Knapsack Problem using Cohort Intelligence Algorithm**

## 4.10 Sankey, Alluvial and Arc Diagrams showing CI Research Insights

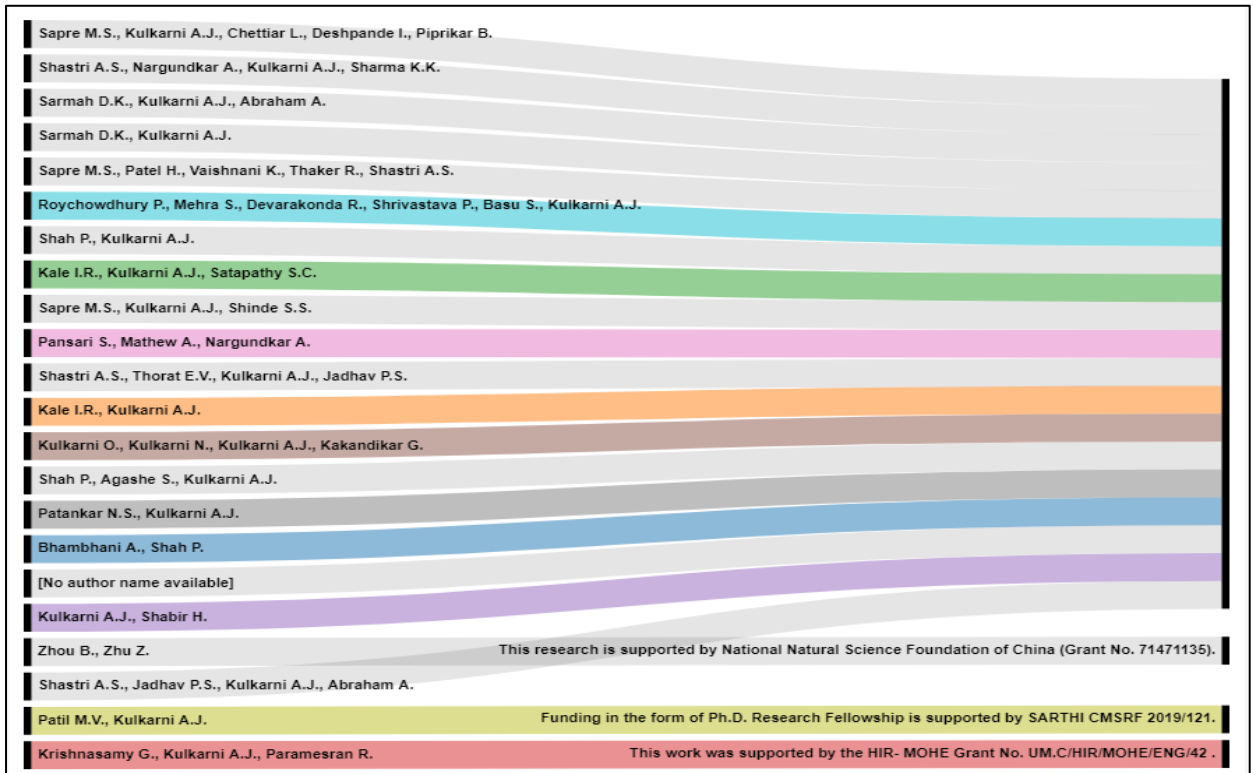

Fig. 33 Authors-Funding-Details Sankey (Scopus_DB)

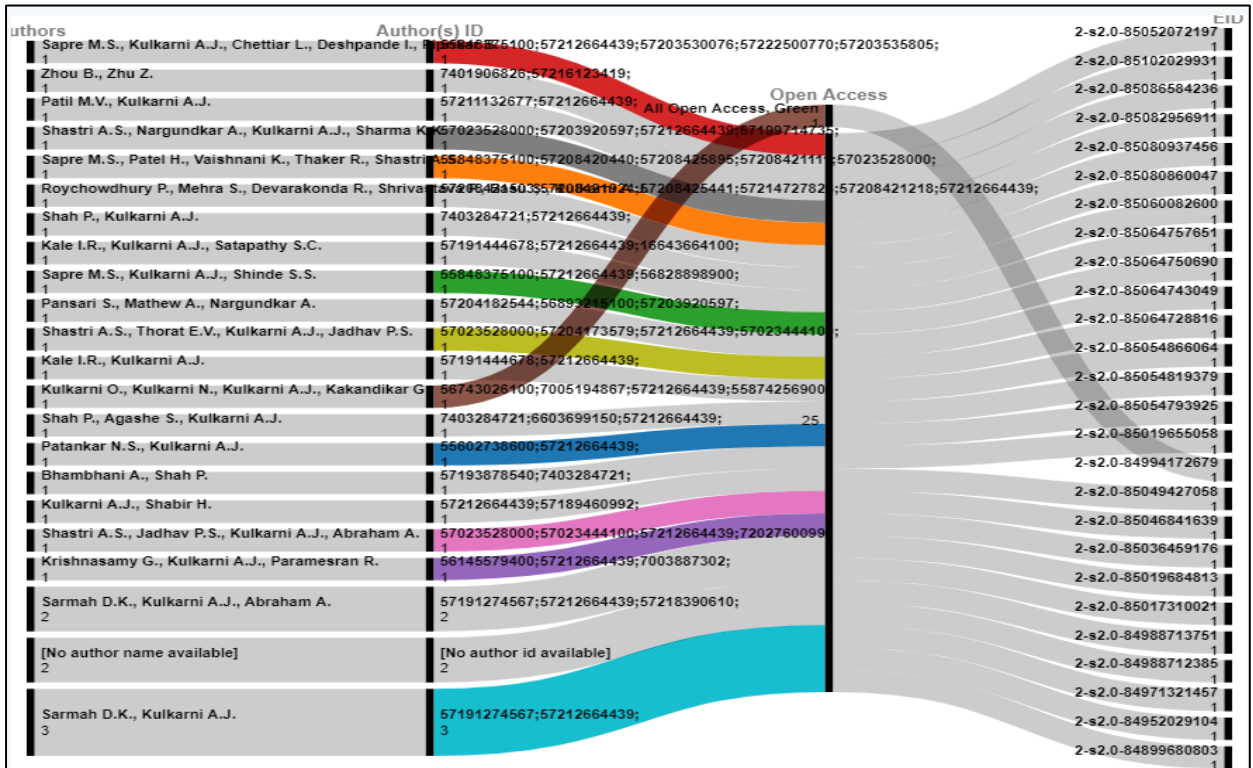

Fig. 34 Authors-Author-ID-EID Alluvial Diagram (Scopus_DB)

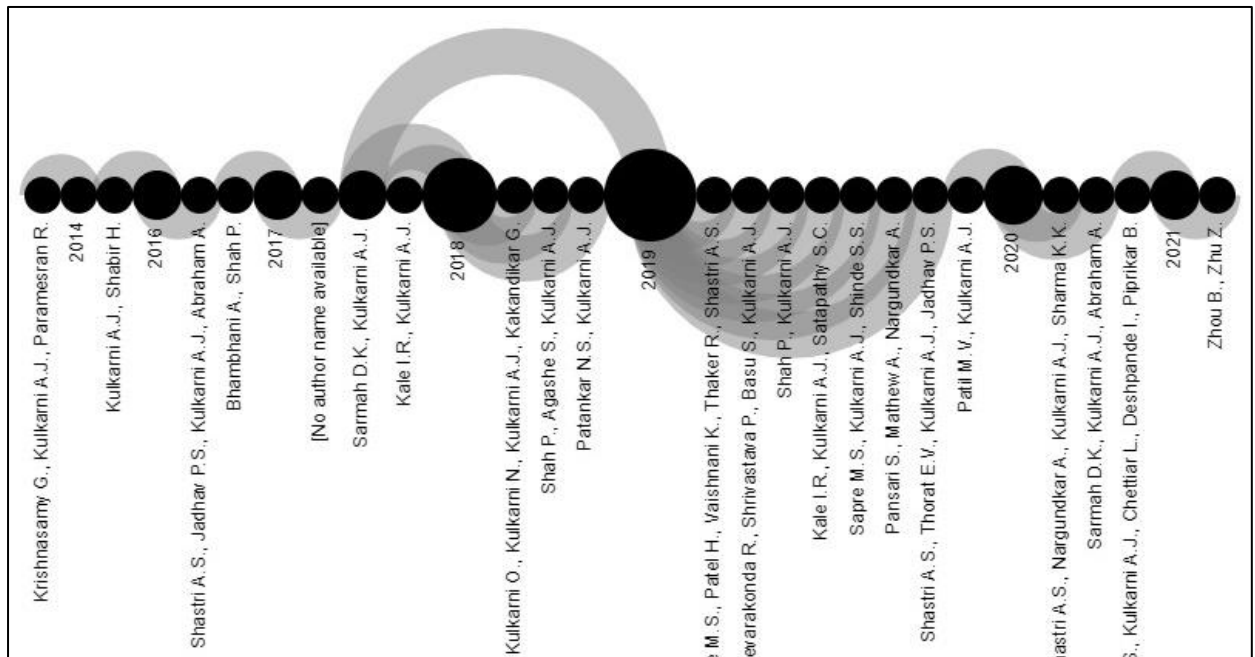

**Fig. 35 Arc Diagram Showing Authors Networking (Scopus_DB)**

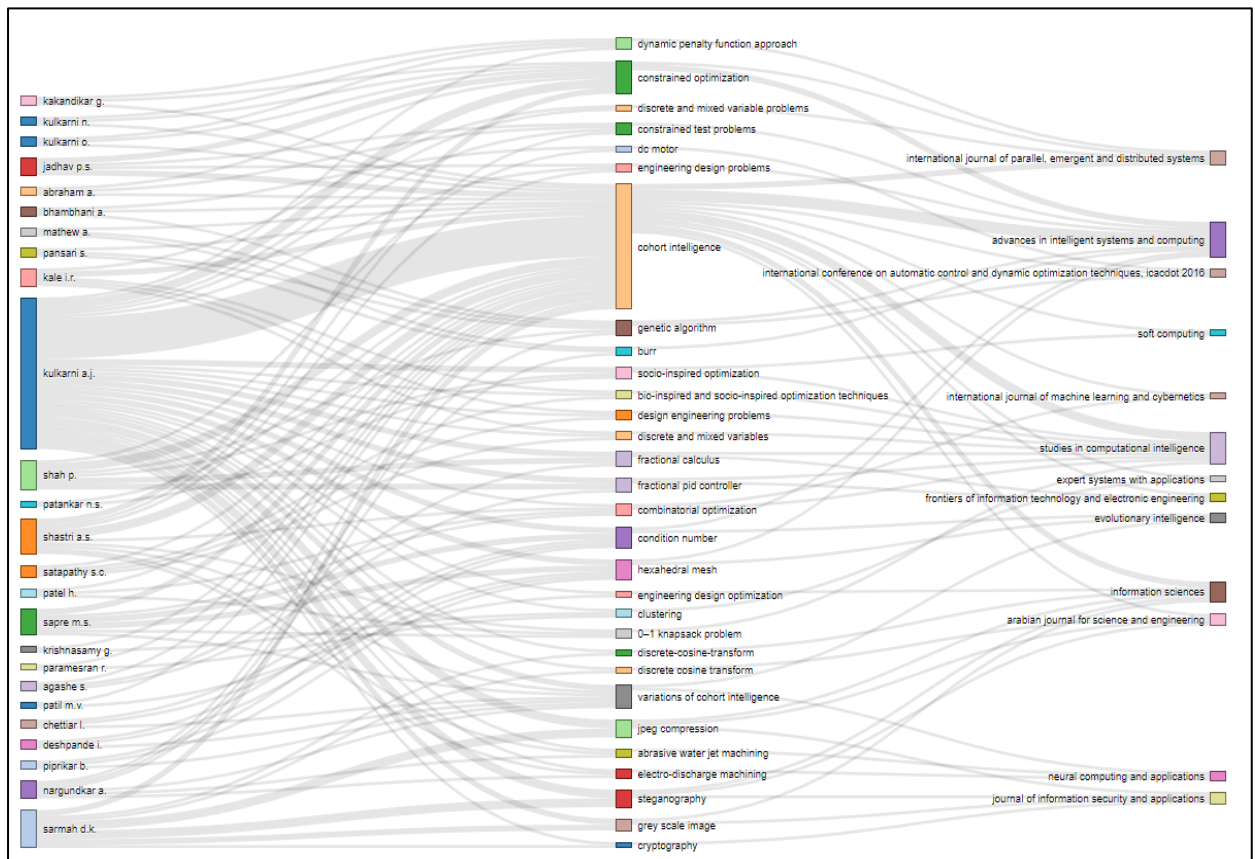

**Fig. 36 A-K-J-Sankey showing Authors-Keywords-Journal's triangulation**

## 5. Concluding Remarks

It is clear from the discussion through sections 2 to 4 that CI is an emerging optimization algorithm. It is observed that CI is applicable to wide number of domains. CI gets evolved over the years. There are only 26 publications till date in the Scopus and 18 in WoS. So, there is a lot of scope for CI related research. There are no review papers. Apart from Computer Science, from the involved authors it also came to notice that CI is prominently utilized in mechanical engineering too. Dr. Anand J Kulkarni who had done pioneering work in CI has posted information related to CI on https://sites.google.com/site/oatresearch/. Apart from small cohort associated with Dr. Anand J Kulkarni, reach of CI need to be widen that will in a way widen its scope too. CI has solved different types of optimization problems successfully. By looking at bibliometric information drafted about CI in this paper, in coming years CI has seemed to get more evolved due to its adaptability, applicability to different subject areas.

**Acknowledgement:** This research article is supported by OAT research lab of Dr. Anand Kulkarni and involved research members of the lab.